\documentstyle[eqsecnum,prd,aps,psfig,floats]{revtex}
\begin{document}
\draft 
\preprint{RAP-312} 
\title{Microwave Background Signatures of a 
        Primordial Stochastic Magnetic Field} 
\author{
Andrew Mack$^{1\,\dagger}$,
Tina Kahniashvili$^{1,2\,\ddagger}$, and
Arthur Kosowsky$^{1,3\,\dagger}$} 
\address{
$^1$Department of Physics and Astronomy,
Rutgers University, 136 Frelinghuysen Road, Piscataway, New Jersey
08854-8019\\ 
$^2$Center for Plasma Astrophysics, 
Abastumani Astrophysical Observatory, A.~Kazbegi Ave.~2a,
380060 Tbilisi, Georgia\\
$^3$School of Natural Sciences, Institute for Advanced Study,
Olden Lane, Princeton, New Jersey 08540\\ 
$^{\dagger}$andymack,\,kosowsky@physics.rutgers.edu,
$^{\ddagger}$tinatin@amorgos.unige.ch}
\maketitle

\begin{abstract}
A stochastic magnetic field in the early Universe will produce
anisotropies in the temperature and polarization of the cosmic
microwave background.  We derive analytic expressions for the
microwave background temperature and polarization power spectra
induced by vector and tensor perturbations from a power-law magnetic
field.  For a scale-invariant stochastic magnetic field smoothed over
a comoving scale of $1\,{\rm Mpc}$, the MAP satellite has the
potential to constrain the comoving mean-field amplitude to be no
greater than approximately $2\times10^{-9}$ G.  Limits improve as
the power-law slope increases: for causally-generated power-law
magnetic fields, the comoving mean-field amplitude has an upper bound
of approximately $4\times10^{-13}\,{\rm G}$.  Such constraints will
surpass all current limits on galactic-scale primordial stochastic
magnetic fields at decoupling.
\end{abstract}

\section{Introduction}
\label{sec:intro}

Magnetic fields of $\mu$G strength are ubiquitous in galaxies
\cite{kronberg94} and clusters of galaxies \cite{kim91}.  The origin
of these fields, however, remains an outstanding problem in cosmology.
It is usually postulated that these $\mu$G fields grew either via some
magnetohydrodynamical (MHD) dynamo mechanism
\cite{zeldovich83,parker79} or via adiabatic compression of a
primordial magnetic field during the collapse of a protogalactic
cloud \cite{piddington64,ohki64,kulsrud90}.
A MHD dynamo requires tiny seed magnetic fields of comoving
amplitude 10$^{-20}$ G in conventional CDM-like cosmological models or
even as tiny as 10$^{-30}$ G in a Universe with a non-zero
cosmological constant \cite{davis99} as suggested by recent
measurements of type Ia supernovae \cite{schmidt98,perlmutter99} and the
microwave background power spectrum 
\cite{miller99,hanany00,balbi00,bernardis00,lange01}.
On the other hand, the adiabatic compression scenario requires a far
larger primordial seed field with comoving amplitude of 10$^{-9}$ G to
10$^{-10}$ G.

Persistent questions about the effectiveness of MHD dynamos
\cite{ko89,vainshtein91,kulsrud92,cattaneo94,gruzinov94,blackman96,kulsrud97,subramanian99}
together with the observation of $\mu$G magnetic fields in
high-redshift galaxies \cite{kronberg94} raise the possibility of a significant
primordial magnetic field in galaxies and clusters of galaxies.  The
origin of such a magnetic field remains a mystery.  Essentially all
viable magnetogenesis mechanisms incorporate speculative ideas in
high-energy theory, including (among others) inflation
\cite{turner88,carroll91,garretson92,ratra92,gasperini95a,gasperini95b},
electroweak \cite{vachaspati91,enqvist93} or QCD
\cite{quashnock89,cheng94} phase transitions, charge asymmetry
\cite{dolgov93}, or a ferromagnetic Yang-Mills vacuum state
\cite{enqvist94}.  As the properties of the primordial magnetic field
predicted varies among these mechanisms, future detections of a
primordial magnetic field may aid us in identifying the correct
magnetogenesis mechanism.  On the cosmological front, a primordial
magnetic field may have affected early-Universe processes such as
phase transitions, baryogenesis, and nucleosynthesis (see
\cite{grasso01} for a review). Relic magnetic fields could provide a
direct source of information about these processes.
A primordial magnetic field may also have influenced structure
formation via contributing to density perturbations on galactic scales
\cite{wasserman78,kim96,tsagas97,tsagas00}
and preserving magnetic energy in Alfv\'en
modes on scales below the Silk damping scale during recombination
\cite{jedamzik98,subramanian98a}.
In short, significant primordial magnetic fields 
would impact both cosmology and particle physics.

The presence of a magnetic field in the early Universe affects the
evolution of metric perturbations, and as a result, produces
temperature and polarization anisotropies in the cosmic microwave
background (CMB). High-resolution measurements of the microwave
background provide a clean and model-independent test for
primordial magnetic fields. We demonstrate in this paper
that fields large enough to result in observed fields via
adiabatic compression will likely leave observable
and distinctive fluctuations in the various power spectra
of microwave background temperature and polarization 
fluctuations.

Substantial progress has been made in understanding the effects of a
primordial magnetic field on the CMB.  The Far Infrared Absolute
Spectrophotometer (FIRAS) upper limits on chemical potential $\mu$ and
Compton $y$ distortions in the CMB blackbody constrain the present
strength of the magnetic field with comoving coherence length between
$400\,{\rm pc}$ and $0.6\,{\rm Mpc}$ to be
$B_0<3\times10^{-8}\,{\rm G}$ \cite{jedamzik00}.  The case of a homogeneous
magnetic field has been considered by several authors.  The best
current constraint on the primordial homogeneous magnetic field
strength is $B_0<3.4\times10^{-9}(\Omega_0h^2_{50})^{1/2}$ G ($h_{50}$
is the present Hubble constant in units of 50 km s$^{-1}$ Mpc$^{-1}$),
obtained by doing statistical analysis on the 4-year Cosmic Background
Explorer (COBE) data for temperature patterns of a Bianchi type VII
anisotropic spacetime \cite{barrow97}.  A primordial homogeneous
magnetic field can produce distortions of the CMB acoustic peaks via
fast magnetosonic waves \cite{adams96}; meanwhile, Alfv\'en wave
excitations can amplify vector perturbations and induce additional
correlations in temperature multipole moments \cite{durrer98}.  
It is shown in Ref.~\cite{kosowsky96} that a
primordial homogeneous magnetic field of present strength 10$^{-9}$ G
at decoupling can induce a measurable Faraday rotation in the CMB
polarization of 1$^\circ$ at a frequency of 30 GHz.  Additional CMB
polarization effects arising from a primordial homogeneous magnetic
field via Faraday rotation include a parity-odd cross correlation
between temperature and polarization anisotropies \cite{scannapieco97}
and the depolarization of the original CMB polarization
\cite{harari97}, which leads to a reduction in the damping of
temperature anisotropies on small angular scales.

The case of a stochastic magnetic field is perhaps more realistic,
because such fields are observed within galaxy clusters
\cite{dreher87,perley91,taylor93,ge93} and predicted by all causal
magnetogenesis mechanisms \cite{grasso01}.  Some numerical estimates
of CMB temperature and polarization power spectra from density
perturbations induced by a primordial stochastic magnetic field are
presented in Ref.~\cite{koh00}, whereas corresponding analytic
estimates, though somewhat crude and valid only for temperature
anisotropies on large angular scales, are given in
Ref.~\cite{lemoine95}.  Effects of Alfv\'en waves induced by a
primordial stochastic magnetic field on CMB temperature and B-type
polarization anisotropies are considered in
Refs.~\cite{subramanian98b} and \cite{seshadri01} respectively.
Finally, a primordial stochastic magnetic field also generates
gravitational waves; the resulting tensor CMB temperature power
spectrum is given in Ref.~\cite{durrer00}.

Although a variety of effects of a primordial stochastic magnetic
field on the CMB have been investigated, the results are fragmented
and a systematic approach is lacking.  Besides the temperature power
spectrum from tensor perturbations given in Ref.~\cite{durrer00}, no
other CMB power spectra have been derived.  We consider a
statistically homogeneous and isotropic stochastic magnetic field with
a power-law power spectrum, generated at some early epoch of the
radiation-dominated Universe.  Based on the computational techniques
in Ref.~\cite{durrer00} and the total angular momentum method for
calculating CMB anisotropies introduced by Hu and White \cite{hu97},
we have completed a comprehensive and unified analytic calculation of
all types of CMB power spectra arising from a primordial stochastic
magnetic field.  This paper focuses on the induced vector and tensor
perturbations.  A primordial magnetic field acts as a continuous
source of vorticity until decoupling and gravitational radiation until
matter-radiation equality.  The resulting vector and tensor
perturbations are one of the few cosmological sources of B-type
polarization \cite{kamionkowski97a,seljak97}, along with primordial
tensor perturbations \cite{kamionkowski98} and gravitational lensing
of the CMB \cite{zaldarriaga98}.  Scalar perturbations induce CMB
anisotropies smaller than those from vector and tensor perturbations
and hence will not be considered here (see Sec.~\ref{sec:discuss}).

In Sec.~\ref{sec:spectrum} we derive the power spectrum for a
primordial stochastic magnetic field.  We then project the vector and
tensor pieces from the electromagnetic stress-energy tensor, from
which we calculate their two-point correlation functions and derive
their isotropic spectra.  Details of the derivation of the vector
isotropic spectrum are presented in the Appendix.
Section~\ref{sec:damping} computes the magnetic damping scales of the
induced vector perturbations at decoupling and tensor
perturbations at matter-radiation equality.
In Sec.~\ref{sec:metric} we review the vector and tensor contributions to
the metric tensor and give their corresponding evolution equations.
We obtain solutions to these equations, which can be expressed as
functions of the magnetic-induced isotropic spectra derived in
Sec.~\ref{sec:spectrum}.  Using the total angular momentum method of
Ref.~\cite{hu97}, we compute analytically the CMB power spectra for
temperature in Sec.~\ref{sec:temperature}, polarization in
Sec.~\ref{sec:polarization}, and the temperature-polarization cross
correlation in Sec.~\ref{sec:cross}.
Section~\ref{sec:discuss} concludes with the physical interpretation of
these results and a discussion of current and future limits on
primordial magnetic fields from the microwave background.
For the vector perturbations, the B-type
is slightly larger than the E-type polarization power spectrum,
whereas the E-type polarization and the cross-correlation
power spectra are approximately identical.  For the tensor perturbations,
the polarization power spectra are actually comparable to the
temperature power spectrum for $n>-3/2$, where $n$ is the magnetic field
power-law spectral index.  As we will show in Sec.~\ref{sec:damping},
the tensor perturbations are damped on smaller scales than the
vector perturbations. The magnetic cutoff wavenumber
determines the overall amplitude of the CMB power spectra, so the
tensor-induced CMB anisotropies will be larger
than the vector anisotropies for $n>-3/2$. 
For a scale-invariant stochastic magnetic field smoothed
over a comoving scale of $1\,{\rm Mpc}$,
near-future microwave background temperature measurements will
constrain the  comoving mean-field amplitude to be no greater 
than approximately 
$2\times10^{-9}$ G.  Limits improve as $n$ increases: 
for causally-generated power-law magnetic fields with $n\geq2$,
the comoving mean-field amplitude will soon have an upper bound of
approximately $4\times10^{-13}\,{\rm G}$.  
These will be the strongest current constraints on galactic-scale primordial 
stochastic magnetic fields at decoupling.
Eventually, precision measurements of the microwave background temperature
and polarization will give significantly stronger constraints.

In this paper, we focus on the induced CMB anisotropies for $l\leq500$,
where the analysis is relatively clean, simple, and free from complications
arising from the last-scattering microphysics.
For simplicity, we consider the case of a flat Universe
with a vanishing cosmological constant.  We employ the following
notational conventions: $a$ is the scale factor,
$\eta$ is the conformal time, overdots are
derivatives with respect to $\eta$, and 0 subscripts denote the
present time.  We set $c=1$ and normalize the scale factor to unity
today.  As usual, Greek indices run from 0 to 3 and Latin ones from 1
to 3.  All calculations are done in Fourier space,
unless real-space dependence is indicated explicitly (as in
Sec.~\ref{sec:spectrum}).  
All magnetic field amplitudes are comoving values,
unless an explicit time dependence is displayed.

\section{Magnetic Power Spectrum and Correlation Functions}
\label{sec:spectrum}

Consider a primordial stochastic magnetic field created at some
specific moment during the radiation-dominated epoch.  The energy
density of the magnetic field is treated as a first-order perturbation
to a flat Friedmann-Robertson-Walker (FRW) background cosmology.
In other words, we do not decompose the magnetic field into a large
homogeneous component and a small fluctuating piece
as in most cases in the literature.
Within the linear approximation, the magnetic field evolves as a stiff
source and we discard all MHD fluid back reactions onto the field
itself \cite{durrer00}.  Prior to decoupling, the conductivity of the
primordial plasma is very large \cite{turner88,ahonen96} and for
practical purposes can be assumed infinite.  In the comoving frame,
this implies the ``frozen-in'' condition ${\mathbf E=-v\times B}$,
where ${\mathbf v}$ is the plasma peculiar velocity and ${\mathbf E}$
is the electric field induced by plasma motions.  Infinite
conductivity leads to a vanishing electric field in linear
perturbation theory $(v\ll 1)$ and allows the time evolution of the
magnetic field to decouple from its spatial structure on sufficiently
large scales.  As the Universe expands, magnetic field lines are
simply conformally diluted due to flux conservation: ${\mathbf
B}(\eta,{\mathbf x})={\mathbf B}({\mathbf x})/a^2$.  On small scales,
however, a primordial magnetic field is damped due to photon and neutrino
 viscosities \cite{jedamzik98,subramanian98a}.  As in
Ref.~\cite{durrer00}, we parametrize this damping by introducing a hard
ultraviolet cutoff wavenumber $k_D$ in the magnetic power spectrum.
We will compute the magnetic damping cutoff wavenumbers
$k_D$'s for both vector and tensor perturbations in Sec.~\ref{sec:damping}.

A statistically homogeneous and isotropic magnetic field must have the
two-point correlation function \cite{durrer00,peacock99}
\begin{equation}
\langle B_i({\mathbf k})B^{*}_j({\mathbf k'})\rangle
=(2\pi)^3P_{ij}P(k)\delta({\mathbf k-k'}),
\label{eq:2pt-fcn-form}
\end{equation}
where
\begin{equation}
P_{ij}\equiv\delta_{ij}-\hat{k}_i\hat{k}_j
\label{eq:proj-operator}
\end{equation}
is a projector onto the transverse plane:
\begin{equation}
P_{ij}P_{jk}=P_{ik},\qquad\qquad P_{ij}\hat{k}_j=0,
\label{eq:proj-properties}
\end{equation}
and $\hat{k}_i=k_i/k$. 
We adopt the Fourier transform convention
\begin{equation}
B_i({\mathbf k}) = \int d^3x\,
\exp(i{\mathbf k\cdot x}) B_i({\mathbf x}).
\label{eq:fourier}
\end{equation}
Note the projection tensor of
Eq.~(\ref{eq:proj-operator}) is valid only for the case of a flat
Universe where perturbations can be decomposed into plane waves; for
non-zero spatial curvatures, the analog to a plane-wave basis must be
employed (see, e.g., \cite{abbott86}).  A specific magnetogenesis model
consists of specifying the function $P(k)$, which we take to be a
power law
\begin{equation}
P(k)=Ak^n.
\label{eq:power-law}
\end{equation}
Our primary interest is to constrain the primordial comoving magnetic field
strength on a certain comoving length scale.
We therefore convolve the field with a 3D-Gaussian
filter transform of comoving radius $\lambda$,
$B_i({\mathbf k})\rightarrow B_i({\mathbf
k})*f_k$, where $f_k=\exp(-\lambda^2 k^2/2)$, and normalize as
\begin{equation}
\langle B_i({\mathbf x})B_i({\mathbf x})\rangle
|_{\lambda}=B^{2}_\lambda.
\label{eq:B-normalization}
\end{equation}
Thus $B_\lambda$ is the magnetic comoving mean-field amplitude obtained by
smoothing over a Gaussian sphere of comoving radius $\lambda$.
The corresponding mean-square value $B^2_\lambda$ is then given by the
Fourier transform of the product of the power spectrum $P(k)$ and the
square of the filter transform $f_k$,
\begin{equation}
B^{2}_\lambda=\frac{2}{(2\pi)^3}\int d^3k\,P(k)|f_k|^2\simeq
\frac{2A}{(2\pi)^2}\frac{1}{\lambda^{n+3}}\Gamma\left(\frac{n+3}{2}\right),
\label{eq:B-meansquare}
\end{equation}
with the factor 2 coming from the trace of the projection 
tensor of Eq.~(\ref{eq:proj-operator}).
We require the spectral index $n>-3$ to prevent infrared divergence of the
integral over the spectrum of long wavelengths $k\rightarrow 0$.
Solving for the normalization constant $A$ and using
Eqs.~(\ref{eq:2pt-fcn-form}) and (\ref{eq:power-law}), we arrive at
the two-point correlation function for a primordial stochastic
magnetic field
\begin{equation}
\langle B_i({\mathbf k})B^{*}_j({\mathbf k'})\rangle
=\frac{(2\pi)^{n+8}}{2}\frac{B^2_\lambda}{\Gamma\left(\frac{n+3}{2}\right)}
P_{ij}\frac{k^n}{k^{n+3}_\lambda}\delta({\mathbf k-k'}),\qquad k<k_D,
\label{eq:2pt-fcn}
\end{equation}
where $k_\lambda=2\pi/{\lambda}$.  The spectrum vanishes for all scales
smaller than the damping scale $k>k_D$.  The
condition $n>-3$ guarantees that superhorizon
coherent fields are not overproduced; 
the limit $n\rightarrow -3$ approaches
a scale-invariant spectrum.
The case $n=0$ corresponds to a white noise spectrum
where we have equal power at all wavelengths.  For a causally-generated 
stochastic magnetic field, we require $n\geq2$ 
\cite{durrer00,peacock99,coles95}.

The induced electromagnetic stress-energy tensor is given by the
convolution of the magnetic field \cite{jackson75}
\begin{equation}
\tau^{(B)}_{ij}({\mathbf k})=\frac{1}{(2\pi)^3}\frac{1}{4\pi} 
\int d^3p\,\left[B_{i}({\mathbf p})B_j({\mathbf k-p})
-\frac{1}{2}\delta_{ij}B_l({\mathbf p})B_l({\mathbf k-p})
\right].
\label{eq:B-stress}
\end{equation}
It can be geometrically decomposed into scalar, vector, and tensor
perturbation modes,
$\tau^{(B)}_{ij}=\Pi_{ij}^{(S)}+\Pi_{ij}^{(V)}+\Pi_{ij}^{(T)}$,
according to their three-space coordinate transformation properties on
the constant-time hypersurface \cite{mukhanov92}.  In the linear
approximation, all types of cosmological perturbations are decoupled
from each other dynamically; thus we can consider each type of
perturbation independently.  From the tensor $\Pi^{(V)}_{ij}$ we can
construct a vector $\Pi^{(V)}_i$ that sources the vorticity
perturbations, whereas the tensor $\Pi^{(T)}_{ij}$ sources the
gravitational wave perturbations.  To obtain CMB power spectra, we
need to derive two-point correlation functions for $\Pi^{(V)}_i$ and
$\Pi^{(T)}_{ij}$ and extract their corresponding isotropic spectra
$|\Pi^{(V),(T)}(k)|^2$.  This is the subject to which we now turn.

\subsection{Vector Projection and Correlation Function}
\label{subsec:vector_proj}

We begin by illustrating how to project from a generic spatial metric
perturbation $\delta g_{ij}$ its vector piece $\delta g^{(V)}_{ij}$.
A vector spatial metric perturbation must have the form
\cite{mukhanov92}
\begin{equation}
\delta g^{(V)}_{ij}=\xi_i\hat{k}_j+\xi_j\hat{k}_i, 
\label{eq:V-form}
\end{equation}
where $\xi_i$ is a divergenceless three vector.
A possible construction for $\xi_i$ is given by
\begin{equation}
\xi_i=\hat{k}_m\delta g_{mi}
-\hat{k}_i\hat{k}_m\hat{k}_n\delta g_{mn}.
\label{eq:V-xi-construct}
\end{equation}
The projection then follows from substituting 
Eq.~(\ref{eq:V-xi-construct}) into Eq.~(\ref{eq:V-form}):
\begin{equation}
\delta g^{(V)}_{ij}=(\hat{k}_m\delta g_{mi}
-\hat{k}_i\hat{k}_m\hat{k}_n\delta g_{mn})\hat{k}_j
+(\hat{k}_m\delta g_{mj}
-\hat{k}_j\hat{k}_m\hat{k}_n\delta g_{mn})\hat{k}_i
=(P_{in}\hat{k}_j+P_{jn}\hat{k}_i)\hat{k}_m\delta g_{mn}.
\label{eq:V-projection}
\end{equation}

Using Eq.~(\ref{eq:V-projection}), the vector part of the
electromagnetic stress-energy tensor is given by
\begin{equation}
\Pi^{(V)}_{ij}=(P_{in}\hat{k}_j+P_{jn}\hat{k}_i)\hat{k}_m\tau^{(B)}_{mn},
\label{eq:B-Vproj}
\end{equation}
from which we can construct a vector $\Pi^{(V)}_i$ via contracting
with the unit vector $\hat{k}_j$,
\begin{equation}
\Pi^{(V)}_i=\Pi^{(V)}_{ij}\hat{k}_j=P_{in}\hat{k}_m\tau^{(B)}_{mn}.
\label{eq:B-vector}
\end{equation}
The physical meaning of $\Pi^{(V)}_i$ is clear upon
examining the Lorentz force vector.  In the infinite conductivity
limit, the Lorentz force vector in real space is given by
\cite{lemoine95,jackson75}
\begin{equation}
{\mathbf L(x)}\simeq -\frac{1}{4\pi}
\left\{{\mathbf B(x)\times\left[\nabla\times B(x)\right]}\right\}
=\frac{1}{4\pi}\left\{\left[{\mathbf B(x)\cdot\nabla}\right]
{\mathbf B(x)}-\frac{1}{2}{\mathbf \nabla}B^2(x)\right\}.
\label{eq:xLorentz-total}
\end{equation}
Fourier transforming Eq.~(\ref{eq:xLorentz-total}), extracting the
corresponding vortical component $L^{(V)}_i$ which satisfies the
divergenceless condition $L^{(V)}_i\hat{k}_i=0$, and comparing with
Eq.~(\ref{eq:B-vector}) shows that
\begin{equation}
L^{(V)}_i=k\Pi^{(V)}_i.
\label{eq:kLorentz-B-vector}
\end{equation}
The vector $\Pi^{(V)}_i$ will appear in the evolution equations for
vector perturbations in Sec.~\ref{subsec:v_perturbations}.  

The stochastic and transverse nature of $\Pi^{(V)}_i$ lead us
to define the two-point correlation function 
\begin{equation}
\langle\Pi^{(V)}_i({\mathbf k})\Pi^{(V)*}_j
({\mathbf k'})\rangle
\equiv P_{ij}|\Pi^{(V)}(k)|^2
\delta({\mathbf k-k'}).
\label{eq:2pt-fcn-vector1} 
\end{equation}
The vector isotropic spectrum $|\Pi^{(V)}(k)|^2$ can be obtained 
using Eq.~(\ref{eq:B-vector}) for $\Pi^{(V)}_i$, evaluating the
two-point correlation function of the electromagnetic stress-energy
tensor of Eq.~(\ref{eq:B-stress}), and comparing the result with 
Eq.~(\ref{eq:2pt-fcn-vector1}).  A lengthy calculation in the Appendix
gives
\begin{equation}
|\Pi^{(V)}(k)|^2\simeq\frac{1}{8\pi(2n+3)}
\left[\frac{(2\pi)^{n+5}B^2_\lambda}
{2\Gamma\left(\frac{n+3}{2}\right)k^{n+3}_\lambda}\right]^2
\left(k^{2n+3}_D+\frac{n}{n+3}k^{2n+3}\right),
\qquad k<k_D.
\label{eq:V-corr-spectrum}
\end{equation}
The first term dominates when $n>-3/2$, whereas the second term
dominates when $-3<n<-3/2$.  For the case $n>-3/2$, 
the vector isotropic spectrum becomes approximately white noise
(independent of $k$) and depends on the ultraviolet cutoff wavenumber $k_D$,
This is because the induced electromagnetic stress-energy tensor of
Eq.~(\ref{eq:B-stress}) is quadratic in the stochastic magnetic field and the
convolution of the magnetic field couples the large and small scale modes.
Each mode of the vector isotropic spectrum is then affected by all scales
of the magnetic power spectrum of Eq.~(\ref{eq:2pt-fcn}) and for the case
of $n>-3/2$, the cutoff scale perturbations completely dominate the
large scale modes (see also Sec.~V of Ref.~\cite{durrer00}).
Note that the term within the square bracket is the normalization
$A$ of the magnetic power spectrum in Eq.~(\ref{eq:power-law}).
To simplify the calculation, 
we will only consider the corresponding dominant term 
for a given spectral index $n$, although including the contributions
from both terms is a straightforward extension of the calculation
presented here. In the neighborhood of $n=-3/2$, both terms must
be included to handle correctly the removable singularity.

\subsection{Tensor Projection and Correlation Function}
\label{subsec:tensor_proj}

Gravitational radiation is produced by the transverse and traceless
piece of the electromagnetic stress-energy tensor, given by
(see, e.g., \cite{durrer00})
\begin{equation}
\Pi^{(T)}_{ij}=(P_{im}P_{jn}-\frac{1}{2}P_{ij}P_{mn})\tau^{(B)}_{mn}.
\label{eq:B-Tproj}
\end{equation}
It follows from the transverse and traceless properties of the tensor
$\Pi^{(T)}_{ij}$ that its two-point correlation function can be
written as \cite{durrer00}
\begin{equation}
\langle\Pi^{(T)}_{ij}({\mathbf k})\Pi^{(T)*}_{lm}
({\mathbf k'})\rangle
\equiv {\mathcal M}_{ijlm}|\Pi^{(T)}(k)|^2
\delta({\mathbf k-k'}).
\label{eq:2pt-fcn-tensor1} 
\end{equation}
The tensor structure ${\mathcal M}_{ijlm}$ is
\begin{eqnarray}
{\mathcal M}_{ijlm} 
&\equiv& P_{il}P_{jm}+P_{im}P_{jl}-P_{ij}P_{lm} \nonumber\\ 
&=& \delta_{il}\delta_{jm}+\delta_{im}\delta_{jl}
-\delta_{ij}\delta_{lm}
+{\hat k}_i{\hat k}_j{\hat k}_l{\hat k}_m \nonumber\\
& & \mbox{}+\delta_{ij}{\hat k}_l{\hat k}_m+\delta_{lm}{\hat k}_i{\hat k}_j
-\delta_{il}{\hat k}_j{\hat k}_m-\delta_{jm}{\hat k}_i{\hat k}_l
-\delta_{im}{\hat k}_j{\hat k}_l-\delta_{jl}{\hat k}_i{\hat k}_m
\label{eq:M_ijlm}
\end{eqnarray}
and satisfies ${\mathcal M}_{ijij}=4$ and 
${\mathcal M}_{iilm} = {\mathcal M}_{ijll} = 0$. 
The tensor isotropic spectrum $|\Pi^{(T)}(k)|^2$ can be obtained using
Eq.~(\ref{eq:B-Tproj}) for $\Pi^{(T)}_{ij}$, evaluating the
two-point correlation function of the electromagnetic stress-energy
tensor of Eq.~(\ref{eq:B-stress}), and comparing the result with 
Eq.~(\ref{eq:2pt-fcn-tensor1}).  A similar calculation as in the case of
the vector isotropic spectrum gives
\begin{equation}
|\Pi^{(T)}(k)|^2\simeq\frac{1}{16\pi(2n+3)}
\left[\frac{(2\pi)^{n+5}B^2_\lambda}
{2\Gamma\left(\frac{n+3}{2}\right)k^{n+3}_\lambda}\right]^2
\left(k^{2n+3}_D+\frac{n}{n+3}k^{2n+3}\right),
\qquad k<k_D.
\label{eq:T-corr-spectrum}
\end{equation}
The tensor isotropic spectrum differs from its vector
counterpart only by a factor of two, due to the ratio of traces of
their corresponding tensor structures $P_{ij}$
and ${\mathcal M}_{ijlm}$.  
Again, the first term dominates when $n>-3/2$, whereas
the second term dominates when $-3<n<-3/2$.  For the case of $n>-3/2$,
the cutoff scale perturbations completely dominate the
large scale modes and hence
the tensor isotropic spectrum depends on the ultraviolet cutoff wavenumber
$k_D$; the resulting tensor CMB temperature power spectrum then
possesses the well-known behavior of a white noise source,
$l^2C_l\propto l^3$ \cite{durrer00}.  We will demonstrate that this
is also true for tensor CMB polarization and temperature-polarization
cross correlation.  As above, the term within the square bracket is the
normalization $A$ of the magnetic power spectrum in Eq.~(\ref{eq:power-law}).
As with the vector case, we will only consider the
dominant term of Eq.~(\ref{eq:T-corr-spectrum}) for a given spectral index $n$.

\section{Magnetic Damping Scales}
\label{sec:damping}

The evolution and damping of primordial magnetic fields are studied
in Refs.~\cite{jedamzik98,subramanian98a}.  These authors consider cases
for which either the magnetic field is linearized about a constant background
field \cite{jedamzik98} or a magnetic field with a tangled component of
unrestricted amplitude is superposed perpendicularly on a homogeneous
field \cite{subramanian98a}.  We are interested in a stochastic
magnetic field with a power-law power spectrum. 
We will first briefly recapitulate the findings of
Refs.~\cite{jedamzik98,subramanian98a}.  Based on these results,
we then proceed to compute the magnetic damping scales separately
for vector and tensor perturbations for a power-law magnetic field.

Primordial magnetic fields are damped by radiative viscosity, which arises
from the finite mean free paths of neutrinos and photons.  Damping of MHD
modes by neutrino viscosity is the most efficient around neutrino decoupling
$(T\sim1\,{\rm MeV})$.  At that time, the neutrino physical mean free path
$(l_{\nu\,\text{dec}}\approx10^{11}\,{\rm cm})$ and the Hubble length
$(H^{-1}_{\nu\,\text{dec}}\approx5\times10^{10}\,{\rm cm})$ are comparable,
hence the dissipation of magnetic energy can only occur on relatively small
scales.  Photon viscosity, on the other hand, damps MHD modes from after
$e^{+}e^{-}$ annihilation $(T\sim20\,{\rm keV})$ until recombination
$(T\sim0.25\,{\rm eV})$; thus it is capable of dissipating magnetic energy
on larger scales.  There are three types of propagating MHD modes: fast and
slow magnetosonic waves, with the fluid velocity making an arbitrary angle
with the background magnetic field; and Alfv\'en waves, with the fluid
velocity oriented perpendicular to the wave vector ${\mathbf k}$ and the
background magnetic field.
Alfv\'en waves induce neither density nor temperature
perturbations.  Fast magnetosonic waves are similar in nature to sound
waves.  Like the acoustically oscillating density fluctuations, they are
Silk-damped by radiation diffusion on scales below the radiation
diffusion length.  Meanwhile, slow magnetosonic and
Alfv\'en waves possess similar behaviors.  During the radiation diffusion
regime $(k^{-1}_{\text{phys}}>l_{\nu,\gamma})$, these waves either
oscillate negligibly or become overdamped, hence the dissipation of magnetic
energy becomes inefficient.  It is only upon entering the free-streaming
regime $(k^{-1}_{\text{phys}}<l_{\nu,\gamma})$ before recombination that
these waves will suffer additional damping.  The resulting maximum damping
scale for these waves is dependent on the background magnetic field strength
and is on the order of the Alfv\'en velocity times the comoving Silk
scale.  Since Alfv\'en modes describe incompressible motions,
we can obtain the magnetic damping scales for vector and tensor perturbations
via computing the damping scales of such modes for a power-law magnetic
power spectrum.

\subsection{Vector Perturbations}
\label{subsec:v_damping}

Since vector perturbations induce CMB anisotropies via vorticity at
recombination, we need to evaluate the Alfv\'en wave damping scales at
recombination for a power-law magnetic field.  Around recombination,
all Alfv\'en modes are overdamped.  The modes that suffer the most
damping while overdamped are those in the free-streaming regime
\cite{jedamzik98,subramanian98a}.  For a non-linear Alfv\'en mode propagating
in a uniform background field $\bar{\mathbf B}$, its free-streaming damping
scale at recombination is given by Eq.~(8.11) of Ref.~\cite{subramanian98a}:
\begin{equation}
k^{-1}_D=\frac{\lambda_D}{2\pi}\approx\sqrt{\frac{3}{5}}V_AL_S
\approx 5.7\times10^{-3}\left(\frac{\bar{B}}
{10^{-9}\,{\rm G}}\right)h^{-1/2}\,{\rm Mpc},
\label{eq:V-damping-scale}
\end{equation}
where $L_S$ is the comoving Silk scale at recombination and we have
assumed $T_{\text{dec}}=0.25\,{\rm eV}$, $\Omega_bh^2=0.0125$, and a
matter-dominated Universe at recombination.  The Alfv\'en velocity $V_A$
arises from the uniform background field $\bar{\mathbf B}$.  For a
linearized Alfv\'en mode, we have to replace $V_A$ by $V_A\cos\theta$,
where $\theta$ is the angle between the wave vector and the zero-order
background field (cf. Eq.~(108) of Ref.~\cite{jedamzik98}).

For a stochastic magnetic field with a power-law power spectrum, the effective
homogeneous magnetic field responsible for the Alfv\'en velocity can be
obtained via smoothing the stochastic field.  As in Ref.~\cite{seshadri01},
we assume the field smoothed over the damping scale $\lambda_D$ acts as the
effective homogeneous field $\bar{B}_{\text{eff}}$.
For each spectral index $n$,
$B^2_\lambda\propto\lambda^{-(n+3)}$ [cf. Eq.~(\ref{eq:B-meansquare})],
and $\bar{B}_{\text{eff}}$ is related to $B_\lambda$ through
(see also Eq.~(26) of Ref.~\cite{durrer00})
\begin{equation}
\bar{B}_{\text{eff}}=B_\lambda\left(\frac{k_D}
{k_\lambda}\right)^{\frac{n+3}{2}}.
\label{eq:B-effective}
\end{equation}
Smoothing the stochastic magnetic field on scales larger than the damping
scale will result in a smaller effective homogeneous field, hence a smaller
effective Alfv\'en velocity and a larger momentum cutoff wavenumber $k_D$.
Since for $n>-3/2$ the vorticity source becomes approximately white noise
(independent of $k$) and is $k_D$-dependent
[cf. Eq.~(\ref{eq:V-corr-spectrum})], a larger momentum cutoff wavenumber
$k_D$ will give rise to larger CMB anisotropies in this regime as we will see.
The estimation in Eq.~(\ref{eq:B-effective}) is therefore a conservative one.
Substituting Eq.~(\ref{eq:B-effective}) into (\ref{eq:V-damping-scale}) yields
\begin{equation}
k_D\approx(1.7\times10^2)^{\frac{2}{n+5}}\left(\frac{B_\lambda}
{10^{-9}\,{\rm G}}\right)^{-\frac{2}{n+5}}\left(\frac{k_\lambda}
{1\,{\rm Mpc}^{-1}}\right)^{\frac{n+3}{n+5}}h^{\frac{1}{n+5}}
\,{\rm Mpc}^{-1}.
\label{eq:V-damping}
\end{equation}
Note that for a given spectral index $n$,
$B^2_\lambda/k^{n+3}_\lambda\propto A$, where $A$ is the normalization of the
magnetic power spectrum in Eq.~(\ref{eq:power-law}).

\subsection{Tensor Perturbations}
\label{subsec:t_damping}

Since the sourcing of gravitational radiation after the Universe becomes
matter-dominated is negligible (see Sec.~\ref{subsec:t_perturbations} and
also Ref.~\cite{durrer00}), the relevant tensor
damping scales are the Alfv\'en wave damping scales at equality.  As in
recombination, all Alfv\'en modes are also overdamped around equality, hence
the modes that are undergoing free streaming suffer the most damping.
The situation is clearly depicted in Fig.~1 of Ref.~\cite{jedamzik98}.
The Alfv\'en wave free-streaming damping scale at equality is
(cf. Eq.~(106) of Ref.~\cite{jedamzik98} and Eq.~(8.10) of
Ref.~\cite{subramanian98a})
\begin{equation}
k^{-1}_D=\frac{\lambda_D}{2\pi}\approx\sqrt{\frac{3}{5}}V_A
L^{\text{diff}}_\gamma(T_{\text{eq}}),
\label{eq:T-damping-scale}
\end{equation}
where $L^{\text{diff}}_\gamma(T_{\text{eq}})$ is the photon comoving
diffusion length at equality,
\begin{eqnarray}
L^{\text{diff}}_\gamma(T_{\text{eq}})
&\approx& 19.5\left(\frac{T_{\text{eq}}}{0.25\,{\rm eV}}\right)^{-5/4}
h^{-1/2}\left(\frac{\Omega_bh^2}{0.0125}\right)^{-1/2}\,{\rm Mpc}
\nonumber\\
&\approx& 0.41h^{-3}\,{\rm Mpc},
\label{eq:photon-diffusion-eq}
\end{eqnarray}
assuming $T_{\text{eq}}=5.5\,{\rm eV}(\Omega_0h^2)$, $\Omega_0=1$, and
$\Omega_bh^2=0.0125$.  Substituting Eq.~(\ref{eq:photon-diffusion-eq})
into (\ref{eq:T-damping-scale}), a similar manipulation as in the case of
vector perturbations gives
\begin{equation}
k_D\approx(8.3\times10^3)^{\frac{2}{n+5}}\left(\frac{B_\lambda}
{10^{-9}\,{\rm G}}\right)^{-\frac{2}{n+5}}\left(\frac{k_\lambda}
{1\,{\rm Mpc}^{-1}}\right)^{\frac{n+3}{n+5}}h^{\frac{6}{n+5}}\,{\rm Mpc}^{-1}.
\label{eq:T-damping}
\end{equation}
Note again that for a given spectral index $n$,
$B^2_\lambda/k^{n+3}_\lambda\propto A$, where $A$ is the normalization of the
magnetic power spectrum in Eq.~(\ref{eq:power-law}).

Several comments are in order.  First, since the tensor source contributes
earlier (at equality) than the vector source (at recombination), tensor
perturbations are damped at smaller scales as illustrated by
Eqs.~(\ref{eq:V-damping}) and (\ref{eq:T-damping}).  For $n>-3/2$, CMB
power spectra are dependent on the momentum cutoff wavenumber $k_D$ and
scale with it as $k^{2n+3}_D$, we therefore expect tensor perturbations
to generate larger anisotropies than the vector perturbations in this regime,
at least for $l\leq500$ that we are considering.  Second, in
Ref.~\cite{durrer00}, the magnetic damping cutoff wavenumber for tensor
perturbations is found to be $4.5\,{\rm Mpc^{-1}}$.  This
value is derived based on the assumption that the Alfv\'en modes are
undergoing damped oscillatory motions.  Our analysis however, shows that
for the magnetic field strengths considered here, the Alfv\'en modes should
be in the overdamped free-streaming regime around equality, 
as also illustrated in Fig.~1 of Ref.~\cite{jedamzik98}.
Finally, as pointed out in Ref.~\cite{jedamzik98},
the Alfv\'en damping scale at equality could in principle be larger
than that given by Eq.~(\ref{eq:T-damping-scale}) since additional
damping could arise due to a possible breakdown of the WKB approximation
in the regime where the Alfv\'en mode is undergoing overdamped free streaming.
In the absence of an accurate quantitative treatment for Alfv\'en damping
scales in this regime, Eq.~(\ref{eq:T-damping-scale}) is our best-educated
guess.  Nevertheless, we caution the readers that with a possible larger
damping scale $k^{-1}_D$ than that given by Eq.~(\ref{eq:T-damping-scale}),
the induced tensor anisotropies for $n>-3/2$ will be reduced accordingly.

\section{Metric Perturbations and Their Evolution}
\label{sec:metric}

A primordial stochastic magnetic field generates CMB anisotropies via
its gravitational effects on the metric tensor.  The full metric
tensor can be decomposed into its background and perturbation pieces,
$g_{\mu\nu}=g^{(0)}_{\mu\nu}+\delta g_{\mu\nu}$; for a flat Universe
with the usual conformal FRW metric,
$g^{(0)}_{\mu\nu}=a^2\eta_{\mu\nu}$, where
$\eta_{\mu\nu}=\mbox{diag}(-1,1,1,1)$ is the Minkowski metric tensor.
The vector (Sec.~\ref{subsec:v_perturbations})
and tensor (Sec.~\ref{subsec:t_perturbations}) perturbations are
calculated separately; scalar perturbations will generally result in smaller
CMB anisotropies compared to vector and tensor contributions, as argued in
Sec.~\ref{sec:discuss}, and so will not be considered here.  We
review the various metric tensor contributions and give the
corresponding evolution equations due to a primordial
stochastic magnetic field.  We then obtain solutions to these equations,
which can be expressed as functions of the isotropic spectra derived
in Sec.~\ref{sec:spectrum}.  

\subsection{Vector Perturbations}
\label{subsec:v_perturbations}

Vector perturbations to the geometry are described by two
divergenceless three-vectors $\zeta_i$ and $\xi_i$ with the general
form [see Eq.~(\ref{eq:V-form}) also]
\begin{equation}
\delta g^{(V)}_{0i}=-a^2\zeta_i,
\qquad\qquad\delta g^{(V)}_{ij}=a^2(\xi_i\hat{k}_j+\xi_j\hat{k}_i).
\label{eq:V-metric-pert}
\end{equation}
Vector perturbations represent vorticity; the divergenceless condition
for vectors $\zeta_i$ and $\xi_i$ guarantees the absence of density
perturbations.  Vector perturbations exhibit gauge freedom, which
arises because the mapping of coordinates between the perturbed
physical manifold and the background is not unique.  From vectors
$\zeta_i$ and $\xi_i$, we can construct a gauge-invariant vector
potential $V_i=\zeta_i+\dot{\xi}_i/k$ that geometrically describes the
vector perturbations of the extrinsic curvature
\cite{bardeen80,durrer94}.  We now exploit the gauge freedom by
explicitly choosing $\xi_i$ to be a constant vector in time; it
follows that $\delta g^{(V)}_{0i}=-a^2V_i$.  Vector perturbations of
the stress-energy tensor can be parametrized by a
divergenceless three-vector ${\mathbf v}^{(V)}$ that perturbs the
four-velocity $u_{\mu}=(1,0,0,0)$ of a stationary fluid element in the
comoving frame \cite{durrer98}:
\begin{equation}
\delta u_{\mu}=(0,{\mathbf v}^{(V)}/a).
\label{eq:pert-4-velocity}
\end{equation}
We can now construct a gauge-invariant, divergenceless three-vector
termed the ``vorticity,''
\begin{equation}
\Omega_i=v^{(V)}_i-V_i.
\label{eq:V-vorticity}
\end{equation}

Two Einstein equations govern vector perturbation evolution.  
The first describes the
vector potential evolution under the influence of a primordial
stochastic magnetic field:
\begin{equation}
\dot{V}_i(\eta,{\mathbf k})+2\frac{\dot{a}}{a}
V_i(\eta,{\mathbf k})=-\frac{16\pi G\Pi^{(V)}_i({\mathbf k})}{a^2k},
\label{eq:V-Einstein-1}
\end{equation}
where $\Pi^{(V)}_i({\mathbf k})$ is given by Eq.~(\ref{eq:B-vector})
and we neglect the vector anisotropic stress of the plasma, which is
in general negligible. The magnetic field source terms 
$\Pi^{(V)}_i({\mathbf k})$ and
$\Pi^{(T)}_{ij}({\mathbf k})$ are expressed in terms of present
comoving magnetic field amplitudes.  Since both of these terms depend
on the magnetic field quadratically, the explicit time dependence 
of the magnetic stress is given by
$\Pi(\eta,{\mathbf k})=\Pi({\mathbf k})/a^4$.
In the absence of the magnetic source term,
the homogeneous solution of this equation behaves like
$V_i\propto1/a^2$. The complete solution including the magnetic source is
simply
\begin{equation}
V_i(\eta,{\mathbf k})=-\frac{16\pi G\Pi^{(V)}_i({\mathbf k})\eta}{a^2k}.
\label{eq:V-Einstein-1-soln}
\end{equation}
During the radiation-dominated epoch we have $a\propto\eta$; a
magnetic field therefore causes vector perturbations to decay less
rapidly ($1/a$ instead of $1/a^2$) with the Universe's expansion.
The second vector Einstein equation is a constraint that relates the
vector potential to the vorticity:
\begin{equation}
-k^2V_i(\eta,{\mathbf k})=16\pi Ga^2(\rho+p)\Omega_i(\eta,{\mathbf k}).
\label{eq:V-Einstein-2}
\end{equation}

Vector conservation equations can be obtained via covariant
conservation of the stress-energy tensor.  Since vector perturbations
cannot generate density perturbations, we have
\begin{equation}
\delta_\gamma=\delta_b=0.
\label{eq:V-energy}
\end{equation}
Before decoupling, photons are coupled to baryons via Thomson
scattering.  The magnetic field affects the photon-baryon fluid
dynamics via the baryons; we therefore introduce the Lorentz force
term into the baryon Euler equation.  The Euler equations for photons
and baryons are respectively \cite{lemoine95,hu97}
\begin{eqnarray}
\dot{\Omega}_{\gamma i}+\dot{\tau}(v^{(V)}_{\gamma i}-v^{(V)}_{bi})
&=& 0,
\label{eq:V-momentum-photon}\\
\dot{\Omega}_{bi}+\frac{\dot{a}}{a}\Omega_{bi}-\frac{\dot{\tau}}{R}
(v^{(V)}_{\gamma i}-v^{(V)}_{bi}) 
&=& \frac{L^{(V)}_i({\mathbf k})}{a^4(\rho_b+p_b)}.
\label{eq:V-momentum-baryon}
\end{eqnarray}
In the above, ${\mathbf\Omega}_{\gamma,b}= {\mathbf
v}^{(V)}_{\gamma,b}-{\mathbf V}$ represent vorticities of photons and
baryons; $\dot{\tau}=n_e\sigma_Ta$ is the differential optical depth
where $n_e$ is the free electron density and $\sigma_T$ is the Thomson
cross section;
$R\equiv(\rho_b+p_b)/(\rho_\gamma+p_\gamma)\simeq3\rho_b/4\rho_\gamma$
is the momentum density ratio between baryons and photons; and
$L^{(V)}_i$ is the vortical piece of the Lorentz force given by 
Eq.~(\ref{eq:kLorentz-B-vector}).  Again, we neglect the small effects due
to the vector anisotropic stress of the plasma.  This set of vector
conservation equations are similar to the one that describes Alfv\'en
waves in Ref.~\cite{adams96}.
Equations (\ref{eq:V-Einstein-1}), (\ref{eq:V-Einstein-2}),
(\ref{eq:V-momentum-photon}), and (\ref{eq:V-momentum-baryon}) are not
independent.  Using the definitions of $R$, $L^{(V)}_i$, and the fact
that $(\rho_\gamma+p_\gamma)\propto 1/a^4$, and solving the Euler
equations in the tight-coupling approximation $v^{(V)}_{\gamma
i}\simeq v^{(V)}_{bi}$, we obtain the following approximate solution
for the vorticity:
\begin{equation}
\Omega_i(\eta,{\mathbf k})\simeq
\frac{k\Pi^{(V)}_i({\mathbf k})\eta}{(1+R)(\rho_{\gamma0}+p_{\gamma0})}.
\label{eq:V-vorticity-soln}
\end{equation}
Note that the same result can be obtained using
Eqs.~(\ref{eq:V-Einstein-1-soln}) 
and (\ref{eq:V-Einstein-2}).  The factor $1+R$ represents reduction in the
vorticity due to the Compton drag of baryons.  At decoupling, the momentum
density ratio between baryons and photons has an approximate value of
$R_{\text{dec}}\simeq3\rho_{b0}/4\rho_{\gamma0}z_{\text{dec}}\simeq0.35$,
where we have assumed $z_{\text{dec}}=1100$ and $\Omega_b h^2=0.0125$.
The vorticity solution of Eq.~(\ref{eq:V-vorticity-soln})
is valid for perturbation
wavelengths larger than the comoving Silk scale $L_S$, where photon viscosity
can be neglected compared to the Lorentz force.
For $k>k_S$, where $k_S=2\pi/L_S$, the Euler equation that includes the
viscous effect of photons is \cite{subramanian98b}
\begin{equation}
\left(\frac{4}{3}\rho_\gamma+\rho_b\right)\dot{\Omega}_i
+\left(\rho_b\frac{\dot{a}}{a}+\frac{k^2\chi}{a}\right)\Omega_i
=\frac{L^{(V)}_i({\mathbf k})}{a^4},
\label{eq:V-momentum-viscosity}
\end{equation}
where $\chi=(4/15)\rho_\gamma L_\gamma a$ is the photon shear viscosity
cofficient and $L_\gamma=\dot{\tau}^{-1}$ is the photon comoving
mean-free path.  In this regime, the vorticity can be obtained 
using the terminal-velocity approximation.  Equating the photon viscosity
term to the Lorentz force,
we obtain \cite{subramanian98b}
\begin{equation}
\Omega_i(\eta,{\mathbf k})\simeq
\frac{\Pi^{(V)}_i(\mathbf k)}{(kL_\gamma/5)(\rho_{\gamma0}+p_{\gamma0})},
\qquad k>k_S.
\label{eq:V-vorticity-viscosity-soln}
\end{equation}

The next step is to
introduce two-point correlation functions for the vector potential and
the vorticity.  Defining their two-point correlation functions as in
Eq.~(\ref{eq:2pt-fcn-vector1}) for the vector $\Pi^{(V)}_i$, and
taking ensemble averages of Eqs.~(\ref{eq:V-Einstein-1-soln}),
(\ref{eq:V-vorticity-soln}), and (\ref{eq:V-vorticity-viscosity-soln}) the
rms isotropic spectra for the vector potential and the vorticity are simply
\begin{equation}
V(\eta,k) = -\frac{16\pi G\Pi^{(V)}(k)\eta}{a^2 k},
\label{eq:V-potential-spectrum}
\end{equation}
\begin{equation}
\Omega(\eta,k)\simeq\cases{\frac{k\Pi^{(V)}(k)\eta}
{(1+R)(\rho_{\gamma0}+p_{\gamma0})}, & $k<k_S$;\cr
\frac{\Pi^{(V)}(k)}{(kL_\gamma/5)(\rho_{\gamma0}+p_{\gamma0})},
				     & $k>k_S$.\cr}
\label{eq:V-vorticity-spectrum}
\end{equation}

Vector perturbations induce CMB temperature
anisotropies via a Doppler and an integrated Sachs-Wolfe effect
\cite{durrer98}:
\begin{equation}
\Theta^{(V)}(\eta_0,{\mathbf k},\hat{{\mathbf n}})
= -{\mathbf v}^{(V)}\cdot\hat{{\mathbf n}}|^{\eta_0}_{\eta_{\text{dec}}}
+ \int^{\eta_0}_{\eta_{\text{dec}}}
d\eta\,\dot{{\mathbf V}}\cdot\hat{{\mathbf n}},
\label{eq:V-CMB-1}
\end{equation}
where $\eta_{\text{dec}}$ represents the conformal time at decoupling.
The decaying nature of the vector potential ${\mathbf V}$ implies that
most of its contributions toward the integrated Sachs-Wolfe term are
around $\eta_{\text{dec}}$.  Neglecting a possible dipole contribution
due to ${\mathbf v}^{(V)}$ today, we obtain \cite{durrer98}
\begin{equation}
\Theta^{(V)}(\eta_0,{\mathbf k},\hat{{\mathbf n}})
\simeq {\mathbf v}^{(V)}(\eta_{\text{dec}},{\mathbf k})
\cdot\hat{{\mathbf n}}-{\mathbf V}(\eta_{\text{dec}},
{\mathbf k})\cdot\hat{{\mathbf n}} 
= {\mathbf\Omega}(\eta_{\text{dec}},{\mathbf k})\cdot\hat{\mathbf n}.
\label{eq:V-CMB-2}
\end{equation}
Vector CMB temperature anisotropies are due to the
vorticity at decoupling.

\subsection{Tensor Perturbations}
\label{subsec:t_perturbations}

Tensor perturbations to the geometry are described by
\begin{equation}
\delta g^{(T)}_{ij}=2a^2h_{ij},
\label{eq:T-metric-pert}
\end{equation}
where $h_{ij}$ is a symmetric, transverse $(h_{ij}\hat{k}_j=0)$, and
traceless $(h_{ii}=0)$ three-tensor.  Unlike vector perturbations,
tensor perturbations have no gauge freedom.

The tensor Einstein equation that describes the evolution of
gravitational waves sourced by a stochastic magnetic field is
\begin{equation}
\ddot{h}_{ij}(\eta,{\mathbf k})+2\frac{\dot{a}}{a}
\dot{h}_{ij}(\eta,{\mathbf k})+k^2h_{ij}(\eta,{\mathbf k})
=8\pi G\Pi^{(T)}_{ij}({\mathbf k})/a^2,
\label{eq:T-Einstein}
\end{equation}
where $\Pi^{(T)}_{ij}({\mathbf k})$ is given by Eq.\
(\ref{eq:B-Tproj}) and as in the case of the vector perturbations, we
neglect the tensor anisotropic stress of the plasma, which is in
general negligible.  Gravitational waves induce CMB temperature
anisotropies by causing photons to propagate along perturbed geodesics
\cite{durrer00,durrer94}:
\begin{equation}
\Theta^{(T)}(\eta_0,{\mathbf k},\hat{{\mathbf n}})
\simeq\int^{\eta_0}_{\eta_{\text{dec}}}d\eta\,
\dot{h}_{ij}(\eta,{\mathbf k}){\hat n}_i{\hat n}_j.
\label{eq:T-CMB-geodesics}
\end{equation}
Our task is, therefore, to obtain the solution for $\dot{h}_{ij}$.  To
calculate tensor CMB power spectra, we need to define two-point
correlation functions for $h_{ij}$ and $\dot{h}_{ij}$ as in
Eq.~(\ref{eq:2pt-fcn-tensor1}) for the tensor $\Pi^{(T)}_{ij}$, with
rms isotropic spectra $h$ and $\dot{h}$ respectively.  Solutions to
the homogeneous equation with $\Pi^{(T)}(k)=0$ are easily obtained.
During the radiation-dominated epoch, $a\propto\eta$ and
$h\propto j_0(k\eta)$ or $y_0(k\eta)$, 
while during the matter-dominated epoch, $a\propto\eta^2$ and
$h\propto j_1(k\eta)/k\eta$ or $y_1(k\eta)/k\eta$, 
where $j_l$ and $y_l$ are the usual spherical Bessel functions.
Assuming the primordial stochastic magnetic field is generated at
$\eta_{\text{in}}$, a Green function technique yields the following
inhomogeneous solution for the radiation-dominated epoch:
\begin{equation}
h(\eta,k)=\frac{2\pi
G\Pi^{(T)}(k)z^2_{\text{eq}}\eta^2_{\text{eq}}}{(3-2\sqrt{2})k\eta}
\int^{\eta}_{\eta_{\text{in}}}
d\eta'\,\frac{\sin[k(\eta-\eta')]}{\eta'},
\qquad\eta<\eta_{\text{eq}},
\label{eq:T-rad-inhomo-soln}
\end{equation}
where $\eta_{\text{eq}}$ denotes the conformal time at
matter-radiation equality.  The magnetic source term on the right hand
side of Eq.~(\ref{eq:T-Einstein}) decays more rapidly with $\eta$ in
the matter-dominated epoch than in the radiation-dominated epoch.  An
approximate solution, therefore, can be obtained by matching the
radiation-dominated inhomogeneous solution of 
Eq.~(\ref{eq:T-rad-inhomo-soln}) to the matter-dominated homogeneous
solutions at equality.
Retaining the dominant contribution, we obtain \cite{durrer00}
\begin{equation}
\dot{h}(\eta,k)\simeq4\pi G\eta^2_0z_{\text{eq}}
\ln\left(\frac{z_{\text{in}}}{z_{\text{eq}}}\right)
k\Pi^{(T)}(k)\frac{j_2(k\eta)}{k\eta},
\qquad\eta>\eta_{\text{eq}}.
\label{eq:T-rms-metric-fluctuations}
\end{equation}

\section{Temperature Power Spectra}
\label{sec:temperature}

We employ the total angular momentum representation introduced by Hu
and White \cite{hu97} to compute the CMB power spectra induced by a
primordial stochastic magnetic field.  By combining intrinsic angular
structure with that of the plane-wave spatial dependence, this
representation renders a transparent description of CMB anisotropy
formation as each moment corresponds directly to an observable angular
sky pattern via its integral solution of the Boltzmann equations.
The CMB temperature power spectrum
today is given by Eq.~(56) of Ref.~\cite{hu97}:
\begin{equation}
C^{\Theta\Theta (X)}_l=\frac{4}{\pi}
\int dk\,k^2\frac{\Theta^{(X)}_l(\eta_0,k)}{2l+1}
\frac{\Theta^{(X)*}_l(\eta_0,k)}{2l+1},
\label{eq:temp-power-spectrum}
\end{equation}
where $X$ stands for $V$ or $T$, and $\Theta_l$'s are the temperature
fluctuation $\Delta T/T$ moments.  
Note that Eq.~(\ref{eq:temp-power-spectrum}) is larger than the corresponding
expression in Ref.~\cite{hu97} by a factor of two as we have already taken
into account the fact that both vector and tensor perturbations
stimulate two modes individually, corresponding to $m=\pm 1,\pm 2$
respectively in the notation of Ref.~\cite{hu97}.
Our strategy is to evaluate the
Boltzmann temperature integral solutions to obtain the $\Theta_l$'s
due to the vector and tensor perturbations.  We then substitute them
into Eq.~(\ref{eq:temp-power-spectrum}) to yield the corresponding
CMB temperature fluctuations spectra.  Though the tensor results are
already given in Ref.~\cite{durrer00}, the vector results derived here are
new.

\subsection{Vector Temperature Power Spectra}
\label{subsec:v_temperature}

The Boltzmann temperature integral solution for vector perturbations
is given by Eqs.\ (61) and (74) of Ref.~\cite{hu97}:
\begin{equation}
\frac{\Theta^{(V)}_l(\eta_0,k)}{2l+1}
= \int^{\eta_0}_0d\eta\,e^{-\tau}
\left\{(\dot{\tau}v^{(V)}_b+\dot{V})j^{(1V)}_l[k(\eta_0-\eta)]
+\dot{\tau}P^{(V)}j^{(2V)}_l[k(\eta_0-\eta)]\right\},
\label{eq:V-temp-int-soln-1}
\end{equation}
where
\begin{equation}
P^{(V)}=\frac{\sqrt{3}}{9}\frac{k}{\dot{\tau}}v^{(V)}_b
\simeq\frac{\sqrt{3}}{9}\frac{k}{\dot{\tau}}\Omega
\label{eq:V-pol-source}
\end{equation}
is the vector polarization source that is generated when tight
coupling breaks down on small scales, where the photon diffusion length 
and the perturbation wavelength become comparable.  The approximation in 
Eq.~(\ref{eq:V-pol-source}) is obtained using 
Eq.~(\ref{eq:V-vorticity}) and noting that $\Omega$ dominates $V$ at
decoupling [cf.\ Eqs.~(\ref{eq:V-potential-spectrum}) and
(\ref{eq:V-vorticity-spectrum})] for $k\agt0.006$ Mpc$^{-1}$, resulting in
$V$ contributing negligibly compared to $\Omega$ upon integrating over $k$'s
to obtain the vector temperature power spectrum in 
Eq.~(\ref{eq:temp-power-spectrum}). Unlike scalar perturbations, 
vector perturbations cannot produce compressional modes 
due to the lack of pressure support.  In the usual case of vector 
perturbations in cosmological fluids without a magnetic field, tight-coupling
expansion of photon and baryon Euler equations give 
$v^{(V)}_b\approx V$, resulting in the vector
polarization source being dependent on the vector potential instead 
(see Eq.~(94) of Ref.~\cite{hu97}). 
A primordial stochastic magnetic field thus enhances
vector polarization by sourcing the vorticity.  The vector
temperature radial functions $j^{(1V)}_l$ and $j^{(2V)}_l$, which
describe how distant sources contribute, are given by Eq.~(15) of
Ref.~\cite{hu97}:
\begin{eqnarray}
j^{(1V)}_l(x)&=&\sqrt{\frac{l(l+1)}{2}}\frac{j_l(x)}{x},\nonumber\\
j^{(2V)}_l(x)&=&\sqrt{\frac{3l(l+1)}{2}}{d\over dx}
\left(\frac{j_l(x)}{x}\right).
\label{eq:V-temp-radial-fcns}
\end{eqnarray}
The optical depth between $\eta$ and $\eta_0$
is defined as $\tau(\eta)\equiv\int^{\eta_0}_\eta
d\eta'\,\dot{\tau}(\eta')$, thus $d\tau/d\eta=-\dot{\tau}$.
Integrating Eq.~(\ref{eq:V-temp-int-soln-1}) by parts using
$de^{-\tau}/d\eta=\dot{\tau}e^{-\tau}$ and
$j^{(2V)}_l(x)=\sqrt{3}(j^{(1V)}_l(x))'$
and Eqs.~(\ref{eq:V-vorticity}) and (\ref{eq:V-pol-source}), we obtain
\begin{equation}
\frac{\Theta^{(V)}_l(\eta_0,k)}{2l+1}=
\int^{\eta_0}_0d\eta\,\dot{\tau}e^{-\tau}
\left\{\Omega j^{(1V)}_l[k(\eta_0-\eta)] +\frac{\sqrt{3}}{9}
\frac{k}{\dot{\tau}}(\Omega+3V)j^{(2V)}_l[k(\eta_0-\eta)]\right\}.
\label{eq:V-temp-int-soln-2}
\end{equation}
For the usual vector perturbations without a magnetic field, the term
proportional to $j^{(1V)}_l$ is strongly suppressed since
$v^{(V)}_b\approx V$ at decoupling \cite{hu97} and hence
$\Omega\simeq0$ as mentioned above.  
Here we have a primordial stochastic magnetic field
sourcing $\Omega$; the term proportional to $j^{(2V)}_l$ is then
suppressed relative to the term proportional to $j^{(1V)}_l$ due to
the factor $k/\dot{\tau}$.  Moreover, $j^{(2V)}_l$ has less angular power
compared to $j^{(1V)}_l$ (see Fig.~3 of Ref.~\cite{hu97}).  Thus to simplify
the calculation, we consider only the term proportional to $j^{(1V)}_l$
in computing the vector temperature integral solution. 
Including the small corrections due to the angular dependence of 
polarization coming from the term proportional to $P^{(V)}j^{(2V)}_l$
and also the vector potential will yield an additional contribution of at most
a few percent toward our final estimate of the vector temperature
power spectra.  The combination $\dot{\tau}e^{-\tau}$ is
the conformal visibility function, which represents the probability
that a photon last scattered within $d\eta$ of $\eta$ and hence is
sharply peaked at the decoupling period.  For $l\leq500$, we can approximate
the vector temperature integral solution reasonably well as
\begin{equation}
\frac{\Theta^{(V)}_l(\eta_0,k)}{2l+1}\simeq\sqrt{\frac{l(l+1)}{2}}
\Omega(\eta_{\text{dec}},k)\frac{j_l(k\eta_0)}{k\eta_0},
\label{eq:V-temp-int-soln-3}
\end{equation}
using Eq.~(\ref{eq:V-temp-radial-fcns}) and the fact that
$\eta_0\gg\eta_{\text{dec}}$.  The vector CMB temperature anisotropies
are due to the vorticity at decoupling, as also illustrated by Eq.\
(\ref{eq:V-CMB-2}).  Substituting Eq.~(\ref{eq:V-temp-int-soln-3})
into the CMB temperature power spectrum expression of
Eq.~(\ref{eq:temp-power-spectrum}) and using 
Eqs.~(\ref{eq:V-vorticity-spectrum}) and (\ref{eq:V-corr-spectrum}),
we obtain
\begin{eqnarray}
C^{\Theta\Theta(V)}_l
&=&\frac{(2\pi)^{2n+11}}{4}l(l+1)\frac{v^4_{A\lambda}} 
{\Gamma^2\left(\frac{n+3}{2}\right)(2n+3)}
\frac{(k_D\eta_0)^{2n+3}}{(k_\lambda\eta_0)^{2n+6}}
\left[\frac{\eta^2_{\text{dec}}}{(1+R_{\text{dec}})^2}\int^{k_S}_0dk\,k
+\frac{25}{L^2_{\gamma\,\text{dec}}}\int^{k_D}_{k_S}\frac{dk}{k^3}\right]
\nonumber\\
& &\times\left[1+\frac{n}{n+3}\left(\frac{k}{k_D}\right)^{2n+3}\right]
J^2_{l+1/2}(k\eta_0),
\label{eq:V-temp-power-spectrum-1}
\end{eqnarray}
where we have defined the Alfv\'en velocity as 
$v_{A\lambda}=B_\lambda/[4\pi(\rho_{\gamma0}+p_{\gamma0})]^{1/2}\simeq
3.8\times10^{-4}(B_\lambda/10^{-9}\,{\rm G})$.  Note that
$(2\pi)^{2n+10}v^4_{A\lambda}/\left[\Gamma^2\left(\frac{n+3}{2}\right)
k^{2n+6}_\lambda\right]\propto A^2$, where $A$ is the normalization of the
magnetic power spectrum in Eq.~(\ref{eq:power-law}).

Depending on whether $n>-3/2$ or $-3<n<-3/2$, we retain only the
corresponding dominant term of the vector isotropic spectrum 
in Eq.~(\ref{eq:V-temp-power-spectrum-1}).  First consider
the case $n>-3/2$, where the vorticity source becomes approximately 
white noise (independent of $k$)
and that $|\Pi^{(V)}(k)|^2$ is dependent on $k_D$.
To obtain an analytic estimate of the integral
$\int^{k_S}_0dk\,kJ^2_{l+1/2}(k\eta_0)$, consider the more
general integral $\int^{x_S}_0dx\,x^pJ^2_l(x)$ for some $p\geq0$.
Since $J^2_l(x)$ only begins to contribute to the integral significantly
when $x\agt l$, in this limit, we employ in the integral the $J_l(x)$
asymptotic expansion for large argument \cite{abramowitz72}:
$J_l(x)\sim\sqrt{2/(\pi x)}\cos[x-(2l+1)\pi/4]$.  Approximating the
oscillations by a factor of one half then gives
\begin{equation}
\int^{x_S}_0dx\,x^pJ^2_l(x)\simeq\int^{x_S}_ldx\,x^pJ^2_l(x)\simeq
\cases{\frac{x^p_S-l^p}{p\pi},                     & $p>0$;\cr
       \frac{1}{\pi}\ln\left(\frac{x_S}{l}\right), & $p=0$.\cr}
\label{eq:V-Bessel-integral}
\end{equation}
The approximation tends to underestimate; it is good to a few percent for
$p>1$ and is within 30\% for $0\leq p\leq 1$.
The integral $\int^{k_S}_0dk\,kJ^2_{l+1/2}(k\eta_0)$ corresponds to
the case $p=1$; the remaining integral on the rhs of
Eq.~(\ref{eq:V-temp-power-spectrum-1})
$\int^{k_D}_{k_S}dk\,k^{-3}J^2_{l+1/2}(k\eta_0)$ can be well-approximated
by $(k^{-3}_S-k^{-3}_D)/(3\pi\eta_0)$, which is good to within 20\%.
As in Eq.~(\ref{eq:V-Bessel-integral}),
this approximation is obtained via employing the Bessel function asymptotic
expansion for large argument, which is justified since $k\eta_0>l$
for $k_S\leq k\leq k_D$ and $l\leq500$.  Keeping only the highest-order
term in $l$, we obtain the vector CMB temperature power spectrum for $n>-3/2$:
\begin{eqnarray}
l^2C^{\Theta\Theta(V)}_l
&=&\frac{(2\pi)^{2n+10}}{2}
\frac{v^4_{A\lambda}l^4}{\Gamma^2\left(\frac{n+3}{2}\right)(2n+3)}
\frac{(k_D\eta_0)^{2n+3}}{(k_\lambda\eta_0)^{2n+6}}
\left\{\left(\frac{\eta_{\text{dec}}/\eta_0}{1+R_{\text{dec}}}\right)^2
(k_S\eta_0-l)\right.
\nonumber\\
& &\left.\mbox{}+\frac{25}{3}\left(\frac{\eta_0}{L_{\gamma\,\text{dec}}}
\right)^2\left[\frac{1}{(k_S\eta_0)^3}-\frac{1}{(k_D\eta_0)^3}\right]\right\},
\qquad n>-3/2.
\label{eq:V-temp-power-spectrum-A}
\end{eqnarray}
The dominant contribution comes from the term proportional to $(k_S\eta_0-l)$,
which arises from the non-damped vorticity for $k<k_S$.  The remaining term
arising from the damped vorticity for $k_S<k<k_D$ gives a negligible
contribution of $<1\%$.  A numerical evaluation of
Eq.~(\ref{eq:V-temp-power-spectrum-1}) shows that its second integral on the
rhs, which arises from the damped vorticity, always contribute negligibly
$(<1\%)$ compared to its first integral for $l\leq500$ and all cases of $n$.
Therefore, we will neglect the damped vorticity contribution when evaluating
the remaining vector CMB temperature power spectra.

For $-3<n<-3/2$, the needed integral is
$\int^{k_S}_0dk\,k^{2n+4}J^2_{l+1/2}(k\eta_0)$. We must
consider three cases depending on whether
the exponent $2n+4$ is greater than, equal to, or less than zero.
For $-2<n<-3/2$,
using Eq.~(\ref{eq:V-Bessel-integral}) for $p=2n+4$, we obtain
\begin{eqnarray}
l^2C^{\Theta\Theta(V)}_l
&=&\frac{(2\pi)^{2n+10}}{4}
\left(\frac{\eta_{\text{dec}}/\eta_0}{1+R_{\text{dec}}}\right)^2
\frac{v^4_{A\lambda} n l^4}
{\Gamma^2\left(\frac{n+3}{2}\right)(2n+3)(n+2)(n+3)(k_\lambda\eta_0)^{2n+6}}
\left[(k_S\eta_0)^{2n+4}-l^{2n+4}\right],
\nonumber\\
& &\qquad -2<n<-3/2.
\label{eq:V-temp-power-spectrum-B}
\end{eqnarray}
For $n=-2$, again using Eq.~(\ref{eq:V-Bessel-integral}) for $p=0$, we obtain
\begin{equation}
l^2C^{\Theta\Theta(V)}_l=2(2\pi)^5
\left(\frac{\eta_{\text{dec}}/\eta_0}{1+R_{\text{dec}}}\right)^2
\frac{v^4_{A\lambda} l^4}{(k_\lambda\eta_0)^2}
\ln\left(\frac{k_S\eta_0}{l}\right),
\qquad n=-2.
\label{eq:V-temp-power-spectrum-C}
\end{equation}
For $-3<n<-2$, numerical evaluation shows that for $-2.3\alt n<-2$, the
integral $\int^{k_S}_0dk\,k^{2n+4}J^2_{l+1/2}(k\eta_0)$ can be
well-approximated by $[(k_S\eta_0)^{2n+4}-l^{2n+4}]/[2\pi(n+2)\eta^{2n+5}_0]$,
which underestimates by at most 30--40\%.  The resulting temperature power
spectrum is then formally identical to that of the case $-2<n<-3/2$:
\begin{eqnarray}
l^2C^{\Theta\Theta(V)}_l
&=&\frac{(2\pi)^{2n+10}}{4}
\left(\frac{\eta_{\text{dec}}/\eta_0}{1+R_{\text{dec}}}\right)^2
\frac{v^4_{A\lambda} n l^4}
{\Gamma^2\left(\frac{n+3}{2}\right)(2n+3)(n+2)(n+3)(k_\lambda\eta_0)^{2n+6}}
\left[(k_S\eta_0)^{2n+4}-l^{2n+4}\right],
\nonumber\\
& &\qquad -2.3\alt n<-2.
\label{eq:V-temp-power-spectrum-D}
\end{eqnarray}
For $-3<n\alt-2.3$, the dominant contribution to the integral
$\int^{k_S}_0dk\,k^{2n+4}J^2_{l+1/2}(k\eta_0)$ is coming from long wavelengths
$k\rightarrow0$, we therefore approximate by integrating over $k$ to infinity.
The resulting integral can be evaluated analytically by using 6.574.2 of 
Ref.~\cite{gradshteyn94},
\begin{equation}
\int^{\infty}_0dk\,J_p(ak)J_q(ak)k^{-b}=
\frac{a^{b-1}\Gamma(b)\Gamma\left(\frac{p+q-b+1}{2}\right)}
{2^b\Gamma\left(\frac{-p+q+b+1}{2}\right)
\Gamma\left(\frac{p+q+b+1}{2}\right)
\Gamma\left(\frac{p-q+b+1}{2}\right)},
\qquad\text{Re}\,(p+q+1)>\text{Re}\,b>0,\,a>0;
\label{eq:GR-6.574.2}
\end{equation}
and 8.335.1 of Ref.~\cite{gradshteyn94},
\begin{equation}
\Gamma(2x)=\frac{2^{2x-1}}{\sqrt{\pi}}\Gamma(x)
\Gamma\left(x+\frac{1}{2}\right).
\label{eq:GR-8.335.1}
\end{equation}
Keeping only the highest-order term in $l$, we finally obtain
\begin{equation}
l^2C^{\Theta\Theta(V)}_l=\frac{(2\pi)^{2n+10}}{2^{2n+7}}
\left(\frac{\eta_{\text{dec}}/\eta_0}{1+R_{\text{dec}}}\right)^2
\frac{\Gamma^2(-n-2)}{\Gamma(-2n-4)\Gamma^2\left(\frac{n+3}{2}\right)}
\frac{v^4_{A\lambda} n l^{2n+8}}{(2n+3)(n+3)(k_\lambda\eta_0)^{2n+6}},
\qquad -3<n\alt-2.3.
\label{eq:V-temp-power-spectrum-E}
\end{equation}
Our approximation overestimates, as expected, and the accuracy improves as
$n$ decreases since more contribution arises from small $k$ and hence
the result will be less sensitive to the upper limit of the integral, which we
have approximated to be infinity.  It is good to within 30\% for
$-2.5\leq n\alt-2.3$ and a few percent for $-3<n<-2.5$.
The temperature power spectrum of each case above has the same $l$ and $k_D$
(with $k_\lambda,k_S\rightarrow k_D$)
dependence as the corresponding spectrum induced by a primordial
homogeneous magnetic field \cite{durrer98} (the correspondence between
the spectral index of Ref.~\cite{durrer98} and ours is $n\rightarrow 2n+3$).
For the sake of completeness, we note that the vector potential
contribution arising from the $j^{(2V)}_l$ term in 
Eq.~(\ref{eq:V-temp-int-soln-2}) will induce temperature power spectra
$l^2C^{\Theta\Theta(V)}_l\propto l^3$ for $n>-3/2$ and 
$l^2C^{\Theta\Theta(V)}_l\propto l^{2n+6}$ for $-3<n<-3/2$.

We now consider the case $n=-3/2$ and show that this apparent singularity is
removable by considering both terms of the vector isotropic spectrum 
in Eq.~(\ref{eq:V-corr-spectrum}).  In the limit $n=-3/2+\varepsilon$, we have
\begin{equation}
|\Pi^{(V)}(k)|^2
\simeq\frac{(2\pi)^6}{16}\frac{B^4_\lambda}{\Gamma^2(3/4)k^3_\lambda}
\frac{1}{2\varepsilon}\left[1-\left(1-\frac{2\varepsilon}{3}\right)
\left(1+\frac{2\varepsilon}{3}\right)^{-1}
\left(\frac{k}{k_D}\right)^{2\varepsilon}\right].
\label{eq:V-corr-spectrum-sing-1}
\end{equation}
Upon expanding the expression within the square bracket to 
$\cal{O}(\varepsilon)$ and using the small-$x$ expansion to the first order,
i.e. $\ln(1+x)\sim x$ for $x\equiv (k-k_D)/k_D$, we obtain
\begin{equation}
|\Pi^{(V)}(k)|^2
\simeq\frac{(2\pi)^6}{16}\frac{B^4_\lambda}{\Gamma^2(3/4)k^3_\lambda}
\left(\frac{5}{3}-\frac{k}{k_D}\right),
\qquad n\approx -3/2.
\label{eq:V-corr-spectrum-sing-2}
\end{equation}
The same result can be obtained via direct substitution of $n=-3/2$ in 
Eqs.~(\ref{eq:V-corr-spectrum2}) to (\ref{eq:k-p-relation}).
Using Eqs.~(\ref{eq:V-corr-spectrum-sing-2}) and (\ref{eq:V-Bessel-integral})
for $p=1\ \text{and}\ 2$, a similar calculation as in 
Eq.~(\ref{eq:V-temp-power-spectrum-A}) gives
\begin{equation}
l^2C^{\Theta\Theta(V)}_l=\frac{(2\pi)^7}{4}
\left(\frac{\eta_{\text{dec}}/\eta_0}{1+R_{\text{dec}}}\right)^2
\frac{v^4_{A\lambda}l^4}{\Gamma^2(3/4)(k_\lambda\eta_0)^3}
\left[\frac{10}{3}(k_S\eta_0-l)-\frac{(k_S\eta_0)^2-l^2}{k_D\eta_0}\right],
\qquad n\approx -3/2,
\label{eq:V-temp-power-spectrum-F}
\end{equation}
thus showing that the singularity at $n=-3/2$ is indeed removable.
For the rest of the paper, we will not produce explicit power spectrum 
expressions for the case $n=-3/2$.  Readers who are interested can easily 
derive the corresponding results via a straightforward extension of the 
calculation outlined above.

\subsection{Tensor Temperature Power Spectra}
\label{subsec:t_temperature}

The Boltzmann temperature integral solution for tensor perturbations
is given by Eqs.~(61) and (74) of Ref.~\cite{hu97}:
\begin{equation}
\frac{\Theta^{(T)}_l(\eta_0,k)}{2l+1}=
\int^{\eta_0}_0d\eta\,e^{-\tau}
[\dot{\tau}P^{(T)}-\dot{h}]
j^{(2T)}_l[k(\eta_0-\eta)],
\label{eq:T-temp-int-soln-1}
\end{equation}
where
\begin{equation}
P^{(T)}=-\frac{1}{3}\frac{\dot{h}}{\dot{\tau}}
\label{eq:T-pol-source}
\end{equation}
is the tensor polarization source and $j^{(2T)}_l$ is the tensor
temperature radial function given by Eq.~(15) of Ref.~\cite{hu97}:
\begin{equation}
j^{(2T)}_l(x)=\sqrt{\frac{3}{8}\frac{(l+2)!}{(l-2)!}}
\frac{j_l(x)}{x^2}.
\label{eq:T-temp-radial-fcn}
\end{equation}
Using Eq.~(\ref{eq:T-rms-metric-fluctuations}) 
and defining $x\equiv k\eta$ and $x_0\equiv k\eta_0$, 
we approximate the tensor temperature integral solution as
(see also Eq.~(18) of Ref.~\cite{durrer00})
\begin{equation}
\frac{\Theta^{(T)}_l(\eta_0,k)}{2l+1}\simeq
-2\pi\sqrt{\frac{8}{3}\frac{(l+2)!}{(l-2)!}}
\left[G\eta^2_0z_{\text{eq}}
\ln\left(\frac{z_{\text{in}}}{z_{\text{eq}}}\right)\right]
\Pi^{(T)}(k)\int^{x_0}_0dx\,
\frac{j_2(x)}{x}\frac{j_l(x_0-x)}{(x_0-x)^2}.
\label{eq:T-temp-int-soln-2}
\end{equation}
The integral above can be numerically approximated as in Eq.~(19) of
Ref.~\cite{durrer00},
\begin{eqnarray}
\int^{x_0}_0dx\,\frac{j_2(x)}{x}\frac{j_l(x_0-x)}{(x_0-x)^2}
&=&\frac{\pi}{2}\int^{x_0}_0dx\, \frac{J_{5/2}(x)}{x^{3/2}}
\frac{J_{l+1/2}(x_0-x)}{(x_0-x)^{5/2}}\nonumber\\
&\simeq&\frac{7\pi}{20}\sqrt{l}\int^{x_0}_0dx\,
\frac{J_{5/2}(x)}{x}\frac{J_{l+1/2}(x_0-x)}{(x_0-x)^3}\nonumber\\
&\simeq&\frac{7\pi}{50}\sqrt{\frac{3l}{2}}\frac{J_{l+3}(x_0)}{x^3_0},
\label{eq:T-Bessel-integral}
\end{eqnarray}
where in going from the second to the third line, we have inserted a 
factor of $\sqrt{3/2}$ for better numerical agreement and used
6.581.2 of Ref.~\cite{gradshteyn94}:
\begin{eqnarray}
\int^a_0dx\,x^{b-1}(a-x)^{-1}J_p(x)J_q(a-x)
&=&\frac{2^b}{aq}\sum^\infty_{m=0}
\frac{(-1)^m\Gamma(b+p+m)\Gamma(b+m)}{m!\Gamma(b)\Gamma(p+m+1)}
(b+p+q+2m)J_{b+p+q+2m}(a),\nonumber\\
& &\qquad\text{Re}\,(b+p)>0,\,\text{Re}\,q>0.
\label{eq:GR-6.581.2}
\end{eqnarray}
Numerical evaluation shows that the approximation in the second line
of Eq.~(\ref{eq:T-Bessel-integral}) is good to 10\% for $l\alt500$.
Substituting Eq.~(\ref{eq:T-Bessel-integral}) into 
Eq.~(\ref{eq:T-temp-int-soln-2}) yields
\begin{equation}
\frac{\Theta^{(T)}_l(\eta_0,k)}{2l+1}\simeq
-\frac{7}{50}(2\pi)^2\sqrt{\frac{l(l+2)!}{(l-2)!}}
\left[G\eta^2_0z_{\text{eq}}\ln\left(\frac{z_{\text{in}}}
{z_{\text{eq}}}\right)\right]
\Pi^{(T)}(k)\frac{J_{l+3}(k\eta_0)}{(k\eta_0)^3}.
\label{eq:T-temp-int-soln-3}
\end{equation}
Using Eqs.~(\ref{eq:T-corr-spectrum}), (\ref{eq:temp-power-spectrum}), 
and (\ref{eq:T-temp-int-soln-3}), we obtain
\begin{eqnarray}
C^{\Theta\Theta(T)}_l
&=&\frac{49}{10000}
(2\pi)^{2n+12}l^2(l-1)(l+1)(l+2)
\left[G\eta^2_0z_{\text{eq}}\ln\left(\frac{z_{\text{in}}}
{z_{\text{eq}}}\right)\right]^2
\frac{B^4_\lambda}{\Gamma^2\left(\frac{n+3}{2}\right)(2n+3)\eta^3_0}
\frac{(k_D\eta_0)^{2n+3}}{(k_\lambda\eta_0)^{2n+6}}\nonumber\\
& &\times\int^{k_D}_0dk\,k^{-4}
\left[1+\frac{n}{n+3}\left(\frac{k}{k_D}\right)^{2n+3}\right]
J^2_{l+3}(k\eta_0).
\label{eq:T-temp-power-spectrum-1}
\end{eqnarray}
Note that $(2\pi)^{2n+10}B^4_\lambda/\left[\Gamma^2\left(\frac{n+3}{2}\right)
k^{2n+6}_\lambda\right]\propto A^2$, where $A$ is the normalization of the
magnetic power spectrum in Eq.~(\ref{eq:power-law}).
 
For $n>-3/2$, the gravitational wave source
is $k_D$-dependent, and the resulting temperature fluctuation spectrum
possesses the well-known behavior $l^2C_l\propto l^3$.  The integral
$\int^{k_D}_0dk\,k^{-4}J^2_{l+3}(k\eta_0)$ can be evaluated using
Eq.~(\ref{eq:GR-6.574.2}); retaining only the
highest-order term in $l$, we obtain
\begin{equation}
l^2C^{\Theta\Theta(T)}_l=\frac{49}{7500}
(2\pi)^{2n+11}\left[G\eta^2_0z_{\text{eq}}
\ln\left(\frac{z_{\text{in}}}{z_{\text{eq}}}\right)\right]^2
\frac{B^4_\lambda l^3}{\Gamma^2\left(\frac{n+3}{2}\right)(2n+3)}
\frac{(k_D\eta_0)^{2n+3}}{(k_\lambda\eta_0)^{2n+6}},
\qquad n>-3/2.
\label{eq:T-temp-power-spectrum-A}
\end{equation}
For $-3<n<-3/2$, a similar calculation gives
\begin{eqnarray}
l^2C^{\Theta\Theta(T)}_l
&=&2^{2n-5}\frac{49}{625}
(2\pi)^{2n+12}\left[G\eta^2_0z_{\text{eq}}
\ln\left(\frac{z_{\text{in}}}{z_{\text{eq}}}\right)\right]^2
\frac{\Gamma(1-2n)}{\Gamma^2(1-n)\Gamma^2\left(\frac{n+3}{2}\right)}
\frac{B^4_\lambda n}{(2n+3)(n+3)}\left(\frac{l}{k_\lambda\eta_0}\right)^{2n+6},
\nonumber\\
& &\qquad -3<n<-3/2.
\label{eq:T-temp-power-spectrum-B}
\end{eqnarray}
Equivalent tensor perturbation results are given in
Eqs.~(20) to (22) of Ref.~\cite{durrer00}.

\section{Polarization Power Spectra}
\label{sec:polarization}

Polarization of the CMB comes in two flavors: E-type and B-type with
electric $(-1)^l$ and magnetic $(-1)^{l+1}$ parities respectively
\cite{zaldarriaga97,kamionkowski97b}.  Physically, they represent
polarization patterns rotated by $\pi/4$ due to the interchanging of Q
and U Stokes parameters.  Vector and tensor perturbations induce both
types of polarizations.  Scalar perturbations, however, cannot
generate B-type polarization due to azimuthal symmetry.  A detection
of the B-type polarization from future high sensitivity CMB
polarization measurements therefore would provide compelling evidence
for vector and/or tensor contributions.  Similar to the CMB
temperature power spectrum of Eq.~(\ref{eq:temp-power-spectrum}), the
E-type and B-type polarization power spectra are
respectively
\begin{eqnarray}
C^{EE(X)}_l&=&\frac{4}{\pi}\int dk\,k^2
\frac{E^{(X)}_l(\eta_0,k)}{2l+1}\frac{E^{(X)*}_l(\eta_0,k)}{2l+1},
\label{eq:E-power-spectrum}\\
C^{BB(X)}_l&=&\frac{4}{\pi}\int dk\,k^2
\frac{B^{(X)}_l(\eta_0,k)}{2l+1}\frac{B^{(X)*}_l(\eta_0,k)}{2l+1},
\label{eq:B-power-spectrum}
\end{eqnarray}
where $X$ stands for $V$ or $T$. 
The correspondence between notations of
Ref.~\cite{hu97} and ours for polarization moments are
$E^{(\pm1)}_l\rightarrow E^{(V)}_l$ and $B^{(\pm1)}_l\rightarrow \pm
B^{(V)}_l$, and similarly for the tensor perturbations.

\subsection{Vector Polarization Power Spectra}
\label{subsec:v_polarization}

\subsubsection{E-type Polarization}
\label{subsubsec:v_E-polarization}

The E-type polarization integral solution for vector perturbations is
\cite{hu97}
\begin{equation}
\frac{E^{(V)}_l(\eta_0,k)}{2l+1}=-\sqrt{6}\int^{\eta_0}_0d\eta\,
\dot{\tau}e^{-\tau}P^{(V)}\epsilon^{(V)}_l[k(\eta_0-\eta)],
\label{eq:V-E-int-soln-1}
\end{equation}
where
\begin{equation}
\epsilon^{(V)}_l(x)=\frac{1}{2}\sqrt{(l-1)(l+2)}
\left[\frac{j_l(x)}{x^2}+\frac{j'_l(x)}{x}\right]
\label{eq:V-E-radial-fcn}
\end{equation}
is the vector E-type polarization radial function given by Eq.~(17) of 
Ref.~\cite{hu97}.  Using Eq.~(\ref{eq:V-pol-source}) for $P^{(V)}$ and the
spherical Bessel function recurrence relation \cite{abramowitz72} 
\begin{equation}
\frac{l}{x}j_l(x)-j'_l(x)=j_{l+1}(x),
\label{eq:AS-10.1.22}
\end{equation}
we approximate the vector E-type polarization integral solution as in
Eq.~(\ref{eq:V-temp-int-soln-3}):
\begin{equation}
\frac{E^{(V)}_l(\eta_0,k)}{2l+1}\simeq-\sqrt{\frac{(l-1)(l+2)}{18}}
kL_{\gamma\,\text{dec}}\Omega(\eta_{\text{dec}},k)\left[(l+1)
\frac{j_l(k\eta_0)}{(k\eta_0)^2}-\frac{j_{l+1}(k\eta_0)}{k\eta_0}\right],
\label{eq:V-E-int-soln-2}
\end{equation}
where $L_{\gamma\,\text{dec}}=\dot{\tau}^{-1}_{\text{dec}}
\simeq 3.39\,{\rm Mpc}$ is the photon comoving mean-free path at decoupling,
assuming $T_{\text{dec}}=0.25\,{\rm eV}$ and $\Omega_bh^2=0.0125$.  For
the tight-coupling approximation to be valid, we require
$kL_{\gamma\,\text{dec}}<1$.
Substituting Eq.~(\ref{eq:V-E-int-soln-2}) into
(\ref{eq:E-power-spectrum}) and using
Eqs.~(\ref{eq:V-vorticity-spectrum}) 
and (\ref{eq:V-corr-spectrum}), we obtain
\begin{eqnarray}
C^{EE(V)}_l
&=&\frac{(2\pi)^{2n+11}}{36}(l-1)(l+2)
\left(\frac{\eta_{\text{dec}}\eta_0}{1+R_{\text{dec}}}\right)^2
\frac{v^4_{A\lambda}}{\Gamma^2\left(\frac{n+3}{2}\right)(2n+3)}
\frac{(k_D\eta_0)^{2n+3}}{(k_\lambda\eta_0)^{2n+6}}
\nonumber\\
& &\times L^2_{\gamma\,\text{dec}}\int^{k_S}_0dk\,k^5
\left[1+\frac{n}{n+3}\left(\frac{k}{k_D}\right)^{2n+3}\right]
\left[(l+1)\frac{J_{l+1/2}(k\eta_0)}{(k\eta_0)^2}
-\frac{J_{l+3/2}(k\eta_0)}{k\eta_0}\right]^2.
\label{eq:V-E-power-spectrum}
\end{eqnarray}
As in the computation of the vector temperature power spectra in
Sec.~\ref{subsec:v_temperature}, we have neglected the damped vorticity
term, which again contributes negligibly $(<3\%)$ for $l\leq500$ and
all cases of $n$.  Note that $(2\pi)^{2n+10}v^4_{A\lambda}/
\left[\Gamma^2\left(\frac{n+3}{2}\right)k^{2n+6}_\lambda\right]\propto A^2$,
where $A$ is the normalization of the magnetic power spectrum in
Eq.~(\ref{eq:power-law}).

Again, depending on whether $n>-3/2$ or $-3<n<-3/2$, we retain only the
corresponding dominant term of the vector isotropic spectrum in
Eq.~(\ref{eq:V-E-power-spectrum}).  A further simplification occurs by
noting that although the cross term proportional to
$J_{l+1/2}(k\eta_0)J_{l+3/2}(k\eta_0)$ is difficult to evaluate analytically,
a numerical evaluation shows that its value is approximately minus
twice that of the term proportional to $J^2_{l+1/2}(k\eta_0)$
for all cases of $n$.  First consider the case $n>-3/2$, where the
vorticity source becomes approximately white noise and is $k_D$-dependent.
The relevant integrals are
\begin{equation}
(l+1)^2\int^{x_S}_0dx\,x J^2_{l+1/2}(x)
\qquad\text{and}\qquad
\int^{x_S}_0dx\,x^3 J^2_{l+3/2}(x),
\label{eq:v_needed_integral1}
\end{equation}
where we have defined $x\equiv k\eta_0$ and $x_S\equiv k_S\eta_0$.  Using
Eq.~(\ref{eq:V-Bessel-integral}) for $p=1$ and 3 respectively for these two
integrals, we obtain
\begin{eqnarray}
l^2C^{EE(V)}_l
&=&\frac{(2\pi)^{2n+10}}{18}
\left(\frac{\eta_{\text{dec}}/\eta_0}{1+R_{\text{dec}}}\right)^2
\left(\frac{L_{\gamma\,\text{dec}}}{\eta_0}\right)^2
\frac{v^4_{A\lambda}l^4}{\Gamma^2\left(\frac{n+3}{2}\right)(2n+3)}
\frac{(k_D\eta_0)^{2n+3}}{(k_\lambda\eta_0)^{2n+6}}
\left[\frac{(k_S\eta_0)^3-l^3}{3}-l^2(k_S\eta_0-l)\right],
\nonumber\\
& &\qquad  n>-3/2.
\label{eq:V-E-power-spectrum-A}
\end{eqnarray}
Comparing to the numerical evaluation of Eq.~(\ref{eq:V-E-power-spectrum})
shows that our approximation is good to a few percent.

For $-3<n<-3/2$, we need to evaluate
\begin{equation}
(l+1)^2\int^{x_S}_0dx\,x^{2n+4}J^2_{l+1/2}(x)
\qquad\text{and}\qquad
\int^{x_S}_0dx\,x^{2n+6}J^2_{l+3/2}(x).
\label{eq:v_needed_integral2}
\end{equation}
Since the exponent within the first integral $2n+4$ changes sign whereas the
exponent within the second integral $2n+6$ remains positive throughout
$-3<n<-3/2$, as in the vector temperature power spectra calculation, we
consider cases depending on whether $2n+4$ is greater than, equal to,
or less than zero in this regime.  For $-2<n<-3/2$, the two integrals of
Eq.~(\ref{eq:v_needed_integral2}) can be approximated using
Eq.~(\ref{eq:V-Bessel-integral}) for $p=2n+4$ and $2n+6$ respectively,
hence
\begin{eqnarray}
l^2C^{EE(V)}_l
&=&\frac{(2\pi)^{2n+10}}{36}
\left(\frac{\eta_{\text{dec}}/\eta_0}{1+R_{\text{dec}}}\right)^2
\left(\frac{L_{\gamma\,\text{dec}}}{\eta_0}\right)^2
\frac{v^4_{A\lambda}nl^4}{\Gamma^2\left(\frac{n+3}{2}\right)
(2n+3)(n+3)(k_\lambda\eta_0)^{2n+6}}
\nonumber\\
& &\times\left[\frac{(k_S\eta_0)^{2n+6}-l^{2n+6}}{n+3}-
l^2\frac{(k_S\eta_0)^{2n+4}-l^{2n+4}}{n+2}\right],
\qquad -2<n<-3/2.
\label{eq:V-E-power-spectrum-B}
\end{eqnarray}
Here our approximation is good to within 10\%.  For $n=-2$, using
Eq.~(\ref{eq:V-Bessel-integral}) for $p=0$ and 2 respectively for
the two integrals of Eq.~(\ref{eq:v_needed_integral2}), we obtain
\begin{equation}
l^2C^{EE(V)}_l=\frac{(2\pi)^5}{9}
\left(\frac{\eta_{\text{dec}}/\eta_0}{1+R_{\text{dec}}}\right)^2
\left(\frac{L_{\gamma\,\text{dec}}}{\eta_0}\right)^2
\frac{v^4_{A\lambda}l^4}{(k_\lambda\eta_0)^2}
\left\{(k_S\eta_0)^2-l^2\left[2\ln
\left(\frac{k_S\eta_0}{l}\right)+1\right]\right\},
\qquad n=-2.
\label{eq:V-E-power-spectrum-C}
\end{equation}
Comparing to the numerical evaluation of Eq.~(\ref{eq:V-E-power-spectrum}),
the approximation here is good to within 10\% in general.  For $-3<n<-2$,
as discussed in Sec.~\ref{subsec:v_temperature}, a numerical evaluation shows
that the first integral of Eq.~(\ref{eq:v_needed_integral2}) can be
well-approximated by $l^2[(k_S\eta_0)^{2n+4}-l^{2n+4}]/(2n+4)\pi$ for
$-2.3\alt n<-2$ [cf. Eq.~(\ref{eq:V-temp-power-spectrum-D})] whereas for
$-3<n\alt-2.3$, it can be approximated using Eq.~(\ref{eq:GR-6.574.2})
since the dominant contribution to the integral arises from long wavelengths
$k\rightarrow0$ [cf. Eq.~(\ref{eq:V-temp-power-spectrum-E})].
The approximation of the second integral of Eq.~(\ref{eq:v_needed_integral2})
using Eq.~(\ref{eq:V-Bessel-integral}) tends to underestimate.
This, however, can be compensated via approximating the first integral of
Eq.~(\ref{eq:v_needed_integral2}) by
$l^2[(k_S\eta_0)^{2n+4}-l^{2n+4}]/(2n+4)\pi$ throughout the regime $-3<n<-2$.  The resulting vector E-type polarization power spectrum is then
formally identical to that of the case $-2<n<-3/2$,
with accuracy good to within 15\% in general.  Hence
\begin{eqnarray}
l^2C^{EE(V)}_l
&=&\frac{(2\pi)^{2n+10}}{36}
\left(\frac{\eta_{\text{dec}}/\eta_0}{1+R_{\text{dec}}}\right)^2
\left(\frac{L_{\gamma\,\text{dec}}}{\eta_0}\right)^2
\frac{v^4_{A\lambda}nl^4}{\Gamma^2\left(\frac{n+3}{2}\right)
(2n+3)(n+3)(k_\lambda\eta_0)^{2n+6}}
\nonumber\\
& &\times\left[\frac{(k_S\eta_0)^{2n+6}-l^{2n+6}}{n+3}-
l^2\frac{(k_S\eta_0)^{2n+4}-l^{2n+4}}{n+2}\right],
\qquad -3<n<-2.
\label{eq:V-E-power-spectrum-D}
\end{eqnarray}

\subsubsection{B-type Polarization}
\label{subsubsec:v_B-polarization}

The B-type polarization integral solution for vector perturbations is
\cite{hu97}
\begin{equation}
\frac{B^{(V)}_l(\eta_0,k)}{2l+1}=-\sqrt{6}\int^{\eta_0}_0d\eta\,
\dot{\tau}e^{-\tau}P^{(V)}\beta^{(V)}_l[k(\eta_0-\eta)],
\label{eq:V-B-int-soln-1}
\end{equation}
where
\begin{equation}
\beta^{(V)}_l(x)=\frac{1}{2}\sqrt{(l-1)(l+2)}\frac{j_l(x)}{x}
\label{eq:V-B-radial-fcn}
\end{equation}
is the vector B-type polarization radial function. 
Using the same approximation in
Eq.~(\ref{eq:V-B-int-soln-1}) as in Eq.~(\ref{eq:V-temp-int-soln-3}), we
obtain
\begin{equation}
\frac{B^{(V)}_l(\eta_0,k)}{2l+1} \simeq -\sqrt{\frac{(l-1)(l+2)}{18}}
kL_{\gamma\,\text{dec}}\Omega(\eta_{\text{dec}},k)
\frac{j_l(k\eta_0)}{k\eta_0},
\label{eq:V-B-int-soln-2}
\end{equation}
which upon substituting into Eq.~(\ref{eq:B-power-spectrum}) and
using Eqs.~(\ref{eq:V-vorticity-spectrum}) and (\ref{eq:V-corr-spectrum}),
yields
\begin{eqnarray}
C^{BB(V)}_l
&=&\frac{(2\pi)^{2n+11}}{36}(l-1)(l+2)
\left(\frac{\eta_{\text{dec}}\eta_0}{1+R_{\text{dec}}}\right)^2
\frac{v^4_{A\lambda}}{\Gamma^2\left(\frac{n+3}{2}\right)(2n+3)}
\frac{(k_D\eta_0)^{2n+3}}{(k_\lambda\eta_0)^{2n+6}}
\nonumber\\
& &\times L^2_{\gamma\,\text{dec}}\int^{k_S}_0dk\,k^5
\left[1+\frac{n}{n+3}\left(\frac{k}{k_D}\right)^{2n+3}\right]
\frac{J^2_{l+1/2}(k\eta_0)}{(k\eta_0)^2},
\label{eq:V-B-power-spectrum}
\end{eqnarray}
where again we have neglected the contribution coming from the damped
vorticity term, which is negligible $(<4\%)$ for $l\leq500$
and all cases of $n$.  Note that $(2\pi)^{2n+10}v^4_{A\lambda}/
\left[\Gamma^2\left(\frac{n+3}{2}\right)k^{2n+6}_\lambda\right]\propto A^2$,
where $A$ is the normalization of the magnetic power spectrum in
Eq.~(\ref{eq:power-law}).
Except for the order of the Bessel function, Eq.~(\ref{eq:V-B-power-spectrum})
is identical to the term proportional to $J^2_{l+3/2}(k\eta_0)$
in the vector E-type polarization power spectrum expression of
Eq.~(\ref{eq:V-E-power-spectrum}).  For $n>-3/2$, using
Eq.~(\ref{eq:V-Bessel-integral}) for $p=3$,
we obtain [cf. Eq.~(\ref{eq:V-E-power-spectrum-A})]
\begin{equation}
l^2C^{BB(V)}_l=\frac{(2\pi)^{2n+10}}{54}
\left(\frac{\eta_{\text{dec}}/\eta_0}{1+R_{\text{dec}}}\right)^2
\left(\frac{L_{\gamma\,\text{dec}}}{\eta_0}\right)^2
\frac{v^4_{A\lambda}l^4}{\Gamma^2\left(\frac{n+3}{2}\right)(2n+3)}
\frac{(k_D\eta_0)^{2n+3}}{(k_\lambda\eta_0)^{2n+6}}
\left[(k_S\eta_0)^3-l^3\right],\qquad  n>-3/2.
\label{eq:V-B-power-spectrum-A}
\end{equation}
For $-3<n<-3/2$, the exponent within the integral $2n+6$ remains positive
throughout; thus using Eq.~(\ref{eq:V-Bessel-integral}) for
$p=2n+6$, we obtain [cf. Eq.~(\ref{eq:V-E-power-spectrum-B})]
\begin{eqnarray}
l^2C^{BB(V)}_l
&=&\frac{(2\pi)^{2n+10}}{36}
\left(\frac{\eta_{\text{dec}}/\eta_0}{1+R_{\text{dec}}}\right)^2
\left(\frac{L_{\gamma\,\text{dec}}}{\eta_0}\right)^2
\frac{v^4_{A\lambda}nl^4}{\Gamma^2\left(\frac{n+3}{2}\right)
(2n+3)(n+3)^2(k_\lambda\eta_0)^{2n+6}}
\left[(k_S\eta_0)^{2n+6}-l^{2n+6}\right],
\nonumber\\
& &\qquad -3<n<-3/2.
\label{eq:V-B-power-spectrum-B}
\end{eqnarray}
Our accuracy here is good to the quality of the analytic approximation in
Eq.~(\ref{eq:V-Bessel-integral}) and is always within 20\%.

\subsection{Tensor Polarization Power Spectra}
\label{subsec:t_polarization}

\subsubsection{E-type Polarization}
\label{subsubsec:t_E-polarization}

The E-type polarization integral solution for tensor perturbations is
\cite{hu97}
\begin{equation}
\frac{E^{(T)}_l(\eta_0,k)}{2l+1}=-\sqrt{6}\int^{\eta_0}_0d\eta\,
\dot{\tau}e^{-\tau}P^{(T)}\epsilon^{(T)}_l[k(\eta_0-\eta)],
\label{eq:T-E-int-soln-1}
\end{equation}
where $P^{(T)}$ is given by Eq.~(\ref{eq:T-pol-source}), and
\begin{equation}
\epsilon^{(T)}_l(x)=\frac{1}{4}\left[-j_l(x)+j''_l(x)
+2\frac{j_l(x)}{x^2}+4\frac{j'_l(x)}{x}\right]
\label{eq:T-E-radial-fcn}
\end{equation}
is the tensor E-type polarization radial function.  
Using Eqs.~(\ref{eq:T-rms-metric-fluctuations}) and (\ref{eq:AS-10.1.22}) and
the spherical Bessel function recurrence relation
\cite{abramowitz72}
\begin{equation}
\frac{l+1}{x}j_l(x)+j'_l(x)=j_{l-1}(x),
\label{eq:AS-10.1.21}
\end{equation}
and defining $x\equiv k\eta$ and $x_0\equiv k\eta_0$, we approximate the
tensor E-type polarization integral solution as
\begin{eqnarray}
\frac{E^{(T)}_l(\eta_0,k)}{2l+1}
&\simeq&\frac{2\pi}{\sqrt{6}}
\left[G\eta^2_0z_{\text{eq}}
\ln\left(\frac{z_{\text{in}}}{z_{\text{eq}}}\right)\right]
\Pi^{(T)}(k)\nonumber\\
&      &\times\int^{x_0}_0dx\,\frac{j_2(x)}{x}
\left\{\left[-2+\frac{(l+1)(l+2)}{(x_0-x)^2}\right]
j_l(x_0-x)-\frac{2}{x_0-x}j_{l+1}(x_0-x)\right\}.
\label{eq:T-E-int-soln-2}
\end{eqnarray}
A similar manipulation as in Eq.~(\ref{eq:T-Bessel-integral}) gives
\begin{equation}
\frac{E^{(T)}_l(\eta_0,k)}{2l+1}\simeq-\frac{7}{100}(2\pi)^2\sqrt{l}
\left[G\eta^2_0z_{\text{eq}}
\ln\left(\frac{z_{\text{in}}}{z_{\text{eq}}}\right)\right]
\frac{\Pi^{(T)}(k)}{k\eta_0}
\left\{\left[1-\frac{l^2}{2(k\eta_0)^2}\right]J_{l+3}(k\eta_0)
+\frac{J_{l+4}(k\eta_0)}{k\eta_0}\right\}.
\label{eq:T-E-int-soln-3}
\end{equation}
Substituting Eq.~(\ref{eq:T-E-int-soln-3}) into
Eq.~(\ref{eq:E-power-spectrum}) and using Eq.~(\ref{eq:T-corr-spectrum}),
we obtain
\begin{eqnarray}
C^{EE(T)}_l
&=&\frac{49}{40000}(2\pi)^{2n+12}l
\left[G\eta^2_0z_{\text{eq}}
\ln\left(\frac{z_{\text{in}}}{z_{\text{eq}}}\right)\right]^2
\frac{B^4_\lambda\eta_0}{\Gamma^2\left(\frac{n+3}{2}\right)(2n+3)}
\frac{(k_D\eta_0)^{2n+3}}{(k_\lambda\eta_0)^{2n+6}}
\nonumber\\
& &\times\int^{k_D}_0dk\,
\left[1+\frac{n}{n+3}\left(\frac{k}{k_D}\right)^{2n+3}\right]
\left\{\left[1-\frac{l^2}{2(k\eta_0)^2}\right]J_{l+3}(k\eta_0)
+\frac{J_{l+4}(k\eta_0)}{k\eta_0}\right\}^2.
\label{eq:T-E-power-spectrum-1}
\end{eqnarray}
Note that $(2\pi)^{2n+10}B^4_\lambda/\left[\Gamma^2\left(\frac{n+3}{2}\right)
k^{2n+6}_\lambda\right]\propto A^2$, where $A$ is the normalization of the
magnetic power spectrum in Eq.~(\ref{eq:power-law}).
For $n>-3/2$, using Eqs.~(\ref{eq:V-Bessel-integral}) and
(\ref{eq:GR-6.574.2}) and keeping only the highest-order terms in $l$ gives
\begin{equation}
l^2C^{EE(T)}_l=\frac{49}{20000}(2\pi)^{2n+11}
\left[G\eta^2_0z_{\text{eq}}
\ln\left(\frac{z_{\text{in}}}{z_{\text{eq}}}\right)\right]^2
\frac{B^4_\lambda l^3}{\Gamma^2\left(\frac{n+3}{2}\right)(2n+3)}
\frac{(k_D\eta_0)^{2n+3}}{(k_\lambda\eta_0)^{2n+6}}
\left[\ln\left(\frac{k_D\eta_0}{l}\right)-\frac{5}{6}\right],
\qquad n>-3/2.
\label{eq:T-E-power-spectrum-A}
\end{equation}
For $-3<n<-3/2$, using Eqs.~(\ref{eq:GR-6.574.2}) and (\ref{eq:GR-8.335.1}) and
keeping only the highest-order terms in $l$, we obtain
\begin{eqnarray}
l^2C^{EE(T)}_l
&=&2^{2n-7}\frac{49}{625}(2\pi)^{2n+12}
\left[G\eta^2_0z_{\text{eq}}
\ln\left(\frac{z_{\text{in}}}{z_{\text{eq}}}\right)\right]^2
\frac{\Gamma(-2n-3)}{\Gamma^2(-n-1)\Gamma^2\left(\frac{n+3}{2}\right)}
\frac{B^4_\lambda(4n^2+3)}{(2n+3)(n+1)(n+3)}
\left(\frac{l}{k_\lambda\eta_0}\right)^{2n+6},
\nonumber\\
& &\qquad -3<n<-3/2.
\label{eq:T-E-power-spectrum-B}
\end{eqnarray}

{}From the properties of radial functions, Hu and White place upper bounds on
how fast various power spectra can grow with $l$ (see Eq.~(78) of
\cite{hu97}).  In particular, tensor polarization power spectra can grow
no faster than $l^2C^{EE,BB(T)}_l\propto l^2$.  Our results for the tensor E-
and B-type (Sec.~\ref{subsubsec:t_B-polarization}) polarization power spectra
seem to violate this constraint for $n>-2$ by an additional factor of $l$,
which arises from numerical approximations as in the second line of
Eq.~(\ref{eq:T-Bessel-integral}).  Within the tensor integral solutions of 
Eqs.~(\ref{eq:T-temp-int-soln-2}), (\ref{eq:T-E-int-soln-2}), and 
(\ref{eq:T-B-int-soln-2}), we have to evaluate integrals of the form
$\int^{x_0}_0dx\,[j_2(x)/x][j_l(x_0-x)/(x_0-x)^p]$.  The piece $j_2(x)/x$
comes from the gravitational wave solution $\dot{h}$ of 
Eq.~(\ref{eq:T-rms-metric-fluctuations}) whereas the piece 
$j_l(x_0-x)/(x_0-x)^p$ comes from the radial functions.  In Ref.~\cite{hu97},
only the radial function properties are used to determine the upper bounds
on the power spectra growth rate, whereas the source behavior has been
entirely neglected.  Our numerical approximation in 
Eq.~(\ref{eq:T-Bessel-integral}) takes into account the source behavior,
i.e. $j_2(x)/x$, and this introduces an additional factor of $l$ in the 
resulting power spectra.  Note that besides the tensor polarization 
power spectra, all the remaining power spectra conform to the growth 
constraints given by Ref.~\cite{hu97}.

\subsubsection{B-type Polarization}
\label{subsubsec:t_B-polarization}

The B-type polarization integral solution for tensor perturbations is
\cite{hu97}
\begin{equation}
\frac{B^{(T)}_l(\eta_0,k)}{2l+1}=-\sqrt{6}\int^{\eta_0}_0d\eta\,\dot{\tau}
e^{-\tau}P^{(T)}\beta^{(T)}_l[k(\eta_0-\eta)],
\label{eq:T-B-int-soln-1}
\end{equation}
where 
\begin{equation}
\beta^{(T)}_l(x)=\frac{1}{2}\left[j'_l(x)+2\frac{j_l(x)}{x}\right]
\label{eq:T-B-radial-fcn}
\end{equation}
is the tensor B-type polarization radial function.  
Using Eqs.~(\ref{eq:T-pol-source}), (\ref{eq:T-rms-metric-fluctuations})
and (\ref{eq:AS-10.1.22}), and defining $x\equiv k\eta$ and
$x_0\equiv k\eta_0$, we approximate the tensor B-type polarization
integral solution as
\begin{equation}
\frac{B^{(T)}_l(\eta_0,k)}{2l+1}\simeq\frac{\sqrt{6}}{3}
(2\pi)\left[G\eta^2_0z_{\text{eq}}
\ln\left(\frac{z_{\text{in}}}{z_{\text{eq}}}\right)\right]
\Pi^{(T)}(k)\int^{x_0}_0dx\,\frac{j_2(x)}{x}
\left[(l+2)\frac{j_l(x_0-x)}{x_0-x}-j_{l+1}(x_0-x)\right].
\label{eq:T-B-int-soln-2}
\end{equation}
A similar manipulation as in Eq.~(\ref{eq:T-Bessel-integral}) gives
\begin{equation}
\frac{B^{(T)}_l(\eta_0,k)}{2l+1}\simeq\frac{7}{100}(2\pi)^2
\sqrt{l}\left[G\eta^2_0z_{\text{eq}}
\ln\left(\frac{z_{\text{in}}}{z_{\text{eq}}}\right)\right]
\frac{\Pi^{(T)}(k)}{k\eta_0}
\left[l\frac{J_{l+3}(k\eta_0)}{k\eta_0}-J_{l+4}(k\eta_0)\right].
\label{eq:T-B-int-soln-3}
\end{equation}
Substituting Eq.~(\ref{eq:T-B-int-soln-3}) into
(\ref{eq:B-power-spectrum}) and using Eq.~(\ref{eq:T-corr-spectrum}),
we obtain
\begin{eqnarray}
C^{BB(T)}_l
&=&\frac{49}{40000}(2\pi)^{2n+12}l
\left[G\eta^2_0z_{\text{eq}}
\ln\left(\frac{z_{\text{in}}}{z_{\text{eq}}}\right)\right]^2
\frac{B^4_\lambda\eta_0}
{\Gamma^2\left(\frac{n+3}{2}\right)(2n+3)}
\frac{(k_D\eta_0)^{2n+3}}{(k_\lambda\eta_0)^{2n+6}}\nonumber\\
& &\times\int^{k_D}_0dk\,
\left[1+\frac{n}{n+3}\left(\frac{k}{k_D}\right)^{2n+3}\right]
\left[l\frac{J_{l+3}(k\eta_0)}{k\eta_0}-J_{l+4}(k\eta_0)\right]^2.
\label{eq:T-B-power-spectrum-1}
\end{eqnarray}
Note that $(2\pi)^{2n+10}B^4_\lambda/\left[\Gamma^2\left(\frac{n+3}{2}\right)
k^{2n+6}_\lambda\right]\propto A^2$, where $A$ is the normalization of the
magnetic power spectrum in Eq.~(\ref{eq:power-law}).
For $n>-3/2$, using Eqs.~(\ref{eq:V-Bessel-integral}) and 
(\ref{eq:GR-6.574.2}), we obtain
\begin{eqnarray}
l^2C^{BB(T)}_l
&=&\frac{49}{20000}(2\pi)^{2n+11}
\left[G\eta^2_0z_{\text{eq}}
\ln\left(\frac{z_{\text{in}}}{z_{\text{eq}}}\right)\right]^2
\frac{B^4_\lambda l^3}{\Gamma^2\left(\frac{n+3}{2}\right)(2n+3)}
\frac{(k_D\eta_0)^{2n+3}}{(k_\lambda\eta_0)^{2n+6}}
\left[\ln\left(\frac{k_D\eta_0}{l}\right)-1\right],
\nonumber\\
& &\qquad n>-3/2.
\label{eq:T-B-power-spectrum-A}
\end{eqnarray}
For $-3<n<-3/2$, a similar calculation gives
\begin{eqnarray}
l^2C^{BB(T)}_l
&=&2^{2n-4}\frac{49}{625}
(2\pi)^{2n+12}\left[G\eta^2_0z_{\text{eq}}
\ln\left(\frac{z_{\text{in}}}{z_{\text{eq}}}\right)\right]^2
\frac{\Gamma(-2n-3)}{\Gamma^2(-n-1)\Gamma^2\left(\frac{n+3}{2}\right)}
\frac{-B^4_\lambda n}{(2n+3)(n+1)(n+3)}
\left(\frac{l}{k_\lambda\eta_0}\right)^{2n+6},
\nonumber\\
& &\qquad -3<n<-3/2.
\label{eq:T-B-power-spectrum-B}
\end{eqnarray}

\section{Cross-Correlation Power Spectra}
\label{sec:cross}

Since temperature $\Theta_l$ has electric parity $(-1)^l$, only $E_l$
couples to $\Theta_l$ in the Thomson scattering and hence $C^{\Theta
E}_l$ is the only possible cross correlation.  The cross-correlation
power spectrum is defined similarly as the temperature and
polarization power spectra:
\begin{equation}
C^{\Theta E(X)}_l=\frac{4}{\pi}\int dk\,k^2
\frac{\Theta^{(X)}_l(\eta_0,k)}{2l+1}\frac{E^{(X)*}_l(\eta_0,k)}{2l+1},
\label{eq:cross-power-spectrum}
\end{equation}
where $X$ stands for $V$ or $T$.

\subsection{Vector Cross-Correlation Power Spectra}
\label{subsec:v_cross}

As discussed in Ref.~\cite{hu97} and shown in its Fig.~5, the vector
dipole radial function $j^{(1V)}_l$ does not correlate well with its
E-type polarization radial function $\epsilon^{(V)}_l$ whereas its quadrupole
radial function $j^{(2V)}_l$ does.  Therefore to compute the vector
cross-correlation power spectra, we need to retain the term proportional to
$j^{(2V)}_l$ in the vector temperature integral solution, though in the 
calculation of the temperature power spectra, we have neglected it since 
it is suppressed relative to the $j^{(1V)}_l$ term.

Beginning with Eq.~(\ref{eq:V-temp-int-soln-2}), retaining the $j^{(2V)}_l$
term, neglecting the vector potential, and using 
Eqs.~(\ref{eq:V-temp-radial-fcns}) and (\ref{eq:AS-10.1.22}), 
we arrive at the following
vector temperature integral solution as in Eq.~(\ref{eq:V-temp-int-soln-3}):
\begin{equation}
\frac{\Theta^{(V)}_l(\eta_0,k)}{2l+1}\simeq
\sqrt{\frac{l(l+1)}{2}}\Omega(\eta_{\text{dec}},k)
\left\{\frac{j_l(k\eta_0)}{k\eta_0}+
\frac{kL_{\gamma\,\text{dec}}}{3}
\left[(l-1)\frac{j_l(k\eta_0)}{(k\eta_0)^2}
-\frac{j_{l+1}(k\eta_0)}{k\eta_0}\right]\right\}.
\label{eq:V-temp-int-soln-4}
\end{equation}
Substituting Eqs.~(\ref{eq:V-temp-int-soln-4}) and (\ref{eq:V-E-int-soln-2})
into (\ref{eq:cross-power-spectrum}) and using 
Eqs.~(\ref{eq:V-vorticity-spectrum}) and (\ref{eq:V-corr-spectrum}), we obtain
\begin{eqnarray}
C^{\Theta E(V)}_l
&=&-\frac{(2\pi)^{2n+11}}{12}\sqrt{l(l-1)(l+1)(l+2)}
\left(\frac{\eta_{\text{dec}}\eta_0}{1+R_{\text{dec}}}\right)^2
\frac{v^4_{A\lambda}}{\Gamma^2\left(\frac{n+3}{2}\right)(2n+3)}
\frac{(k_D\eta_0)^{2n+3}}{(k_\lambda\eta_0)^{2n+6}}\nonumber\\
& &\times L_{\gamma\,\text{dec}}\int^{k_S}_0dk\,k^4
\left[1+\frac{n}{n+3}\left(\frac{k}{k_D}\right)^{2n+3}\right]
\left\{(l+1)\frac{J^2_{l+1/2}(k\eta_0)}{(k\eta_0)^3}
-\frac{J_{l+1/2}(k\eta_0)J_{l+3/2}(k\eta_0)}{(k\eta_0)^2}\right.\nonumber\\
& &\left.\mbox{}+\frac{kL_{\gamma\,\text{dec}}}{3}
\left[(l^2-1)\frac{J^2_{l+1/2}(k\eta_0)}{(k\eta_0)^4}
-2l\frac{J_{l+1/2}(k\eta_0)J_{l+3/2}(k\eta_0)}{(k\eta_0)^3}
+\frac{J^2_{l+3/2}(k\eta_0)}{(k\eta_0)^2}\right]\right\},
\label{eq:V-cross-power-spectrum}
\end{eqnarray}
where again we have neglected the contribution coming from the damped
vorticity term, which is negligible $(<3\%)$ for $l\leq500$ and all
cases of $n$.  Note that $(2\pi)^{2n+10}v^4_{A\lambda}/
\left[\Gamma^2\left(\frac{n+3}{2}\right)k^{2n+6}_\lambda\right]\propto A^2$,
where $A$ is the normalization of the magnetic power spectrum in
Eq.~(\ref{eq:power-law}).
The first two terms within the curly bracket arise from
correlating $j^{(1V)}_l$ with $\epsilon^{(V)}_l$.  Although the second term
proportional to $J_{l+1/2}(k\eta_0)J_{l+3/2}(k\eta_0)$ cannot be approximated
analytically, a numerical evaluation however shows that these two terms
always roughly cancel each other, which agrees with Ref.~\cite{hu97}
that $j^{(1V)}_l$ does not correlate well with $\epsilon^{(V)}_l$.
The remaining three terms arise from correlating $j^{(2V)}_l$
with $\epsilon^{(V)}_l$.  In the limit $l\gg1$, 
these terms and the three Bessel terms within the vector E-type
polarization power spectrum expression of Eq.~(\ref{eq:V-E-power-spectrum})
are almost identical.  To simplify the approximation, we will neglect the
two terms arising from the correlation between $j^{(1V)}_l$
and $\epsilon^{(V)}_l$.  Thus apart from an overall minus sign, the resulting
power spectra for all cases here are approximately equal to the corresponding
vector E-type polarization power spectra, given in
Eqs.~(\ref{eq:V-E-power-spectrum-A}) and (\ref{eq:V-E-power-spectrum-B})
to (\ref{eq:V-E-power-spectrum-D}).

Since we have neglected two terms in the vector cross-correlation power
spectrum expression of Eq.~(\ref{eq:V-cross-power-spectrum}), accuracy here
is worse than that of the corresponding E-type polarization power spectra.
Note that the terms arising from the correlation between
$j^{(2V)}_l$ and $\epsilon^{(V)}_l$ are suppressed by an additional factor
of $kL_{\gamma\,\text{dec}}$ relative to the two terms arising from the
correlation between $j^{(1V)}_l$ and $\epsilon^{(V)}_l$.  Because of
this suppression factor, a numerical calculation shows that the residuals
of the first two neglected terms can easily be the same order as
the remaining retained terms, reducing the accuracy of our approximation.
Our approximation is good to within a factor of three in general
and tends to underestimate.

\subsection{Tensor Cross-Correlation Power Spectra}
\label{subsec:t_cross}

Using Eqs.~(\ref{eq:T-temp-int-soln-3}), (\ref{eq:T-E-int-soln-3}),
and (\ref{eq:cross-power-spectrum}), we obtain the tensor
cross-correlation power spectrum expression
\begin{eqnarray}
C^{\Theta E(T)}_l
&=&\frac{49}{20000}(2\pi)^{2n+12}l^3
\left[G\eta^2_0z_{\text{eq}}
\ln\left(\frac{z_{\text{in}}}{z_{\text{eq}}}\right)\right]^2
\frac{B^4_\lambda}{\Gamma^2\left(\frac{n+3}{2}\right)(2n+3)\eta_0}
\frac{(k_D\eta_0)^{2n+3}}{(k_\lambda\eta_0)^{2n+6}}\nonumber\\
& &\times\int^{k_D}_0dk\,k^{-2}\left[1+\frac{n}{n+3}
\left(\frac{k}{k_D}\right)^{2n+3}\right]
\left\{\left[1-\frac{l^2}{2(k\eta_0)^2}\right]J^2_{l+3}(k\eta_0)
+\frac{J_{l+3}(k\eta_0)J_{l+4}(k\eta_0)}{k\eta_0}\right\}.
\label{eq:T-cross-power-spectrum}
\end{eqnarray}
Note that $(2\pi)^{2n+10}B^4_\lambda/\left[\Gamma^2\left(\frac{n+3}{2}\right)
k^{2n+6}_\lambda\right]\propto A^2$, where $A$ is the normalization of the
magnetic power spectrum in Eq.~(\ref{eq:power-law}).
For $n>-3/2$, using Eq.~(\ref{eq:GR-6.574.2}) and keeping only
the highest-order terms in $l$, we obtain
\begin{equation}
l^2C^{\Theta E(T)}_l=\frac{49}{15000}
(2\pi)^{2n+11}\left[G\eta^2_0z_{\text{eq}}
\ln\left(\frac{z_{\text{in}}}{z_{\text{eq}}}\right)\right]^2
\frac{B^4_\lambda l^3}{\Gamma^2\left(\frac{n+3}{2}\right)(2n+3)}
\frac{(k_D\eta_0)^{2n+3}}{(k_\lambda\eta_0)^{2n+6}},
\qquad n>-3/2.
\label{eq:T-cross-power-spectrum-A}
\end{equation}

For $-3<n<-3/2$, a similar calculation gives
\begin{eqnarray}
l^2C^{\Theta E(T)}_l
&=& 2^{2n-6}\frac{49}{625}(2\pi)^{2n+12}
\left[G\eta^2_0z_{\text{eq}}
\ln\left(\frac{z_{\text{in}}}{z_{\text{eq}}}\right)\right]^2
\frac{\Gamma(-2n-1)}{\Gamma^2(-n)\Gamma^2\left(\frac{n+3}{2}\right)}
\frac{B^4_\lambda(2n-1)}{(2n+3)(n+3)}
\left(\frac{l}{k_\lambda\eta_0}\right)^{2n+6},\nonumber\\
& & \qquad -3<n<-3/2.
\label{eq:T-cross-power-spectrum-B}
\end{eqnarray}

\section{Results and Discussion}
\label{sec:discuss}

\begin{figure}
\psfig{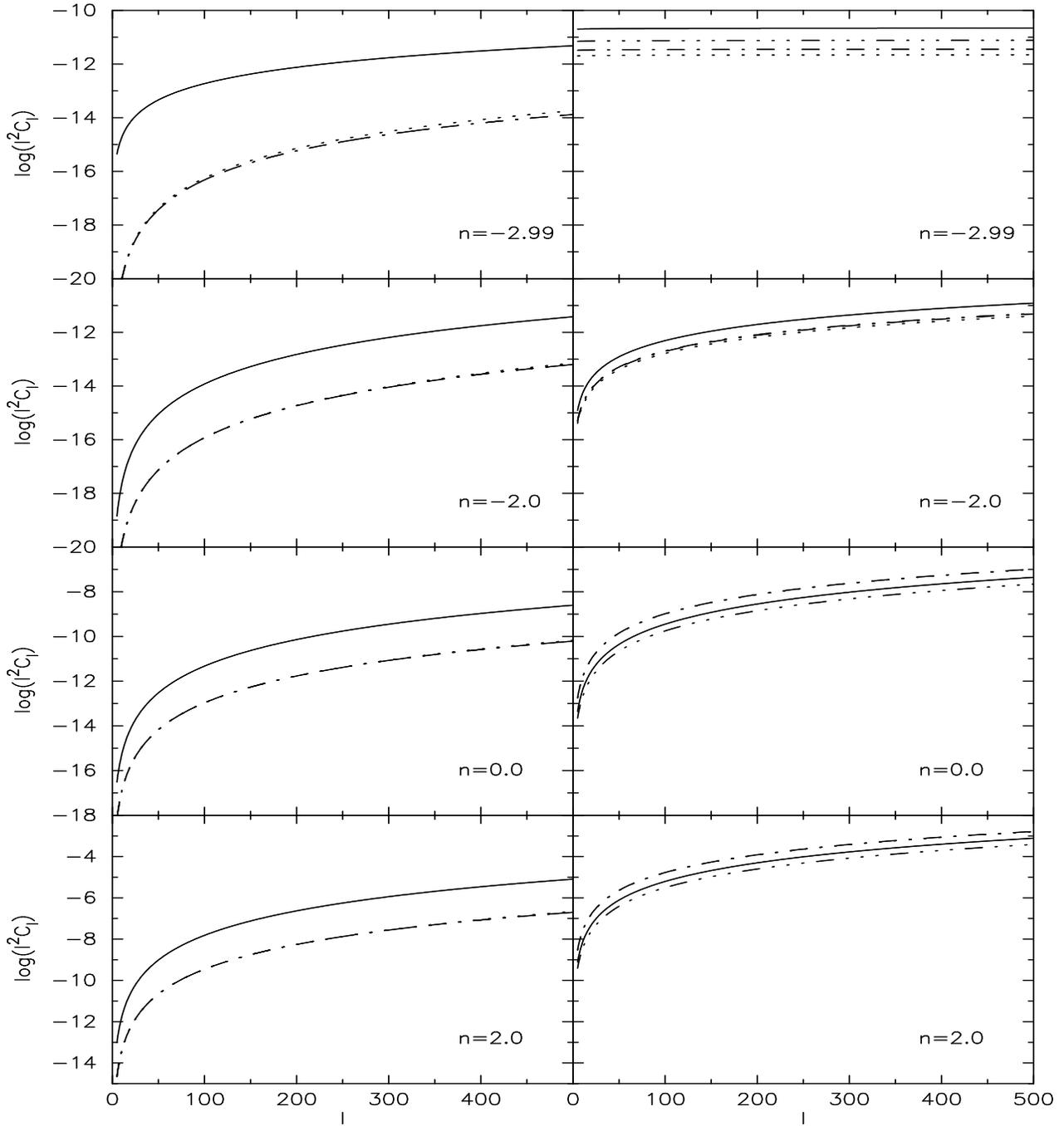}
\vskip 0.5cm
\caption{The microwave background power spectra for vector (left panels)
and tensor (right panels) perturbations from a power-law stochastic
magnetic field with spectral index $n$.  Solid line represents $\Theta\Theta$,
dash-dot line represents EE, dotted line represents BB, and
dash-dot-dot-dot line represents $\Theta\mbox{E}$.  The magnetic comoving
mean-field amplitude is chosen to be $B_\lambda=10^{-9}\,{\rm G}$,
with a smoothing Gaussian sphere comoving radius of $\lambda=1\,{\rm Mpc}$.
The magnetic damping cutoff wavenumbers for vector and tensor perturbations
are given by Eqs.~(\ref{eq:V-damping}) and (\ref{eq:T-damping}) respectively.
The absolute values of the vector cross correlations are plotted.
For the tensor perturbations, we assume $z_{\text{in}}/z_{\text{eq}}=10^9$.}
\label{fig:Figure1}
\end{figure}

\begin{figure}
\psfig{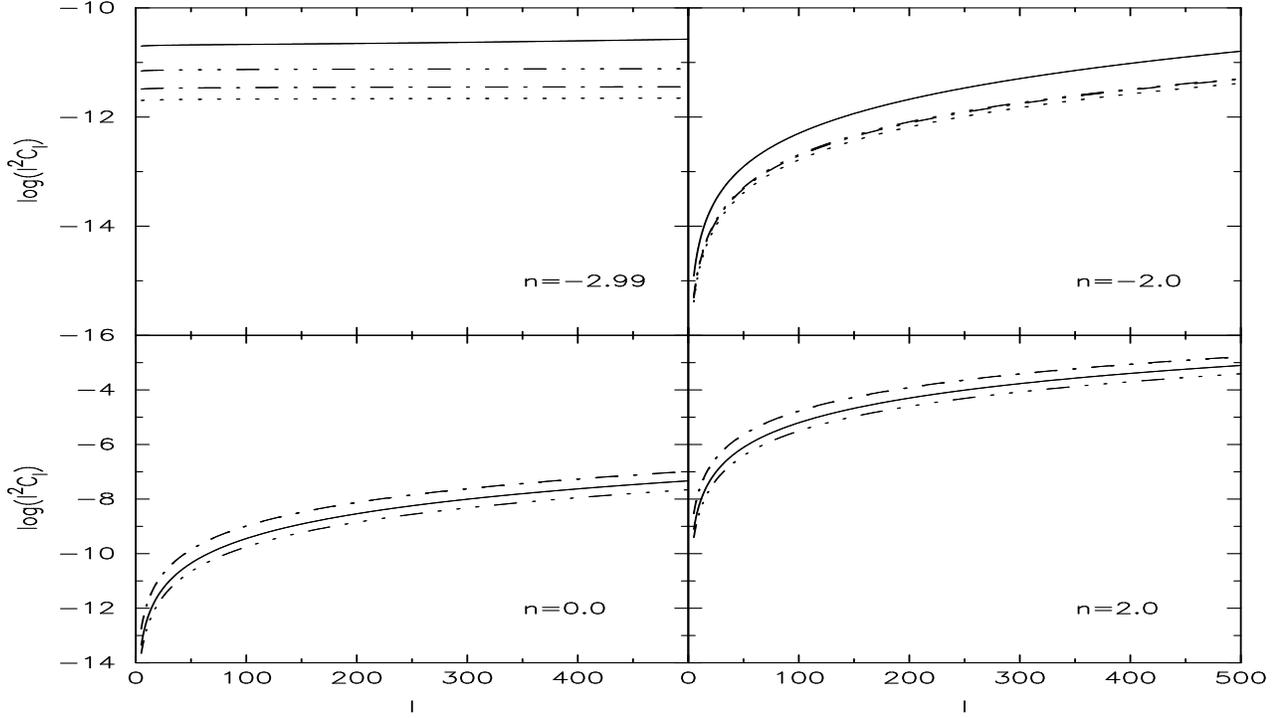}
\vskip 0.5cm
\caption{Same as in Fig.~\ref{fig:Figure1}, except that the microwave
background power spectra are for vector plus tensor perturbations.}
\label{fig:Figure2}
\end{figure}

\begin{figure}
\psfig{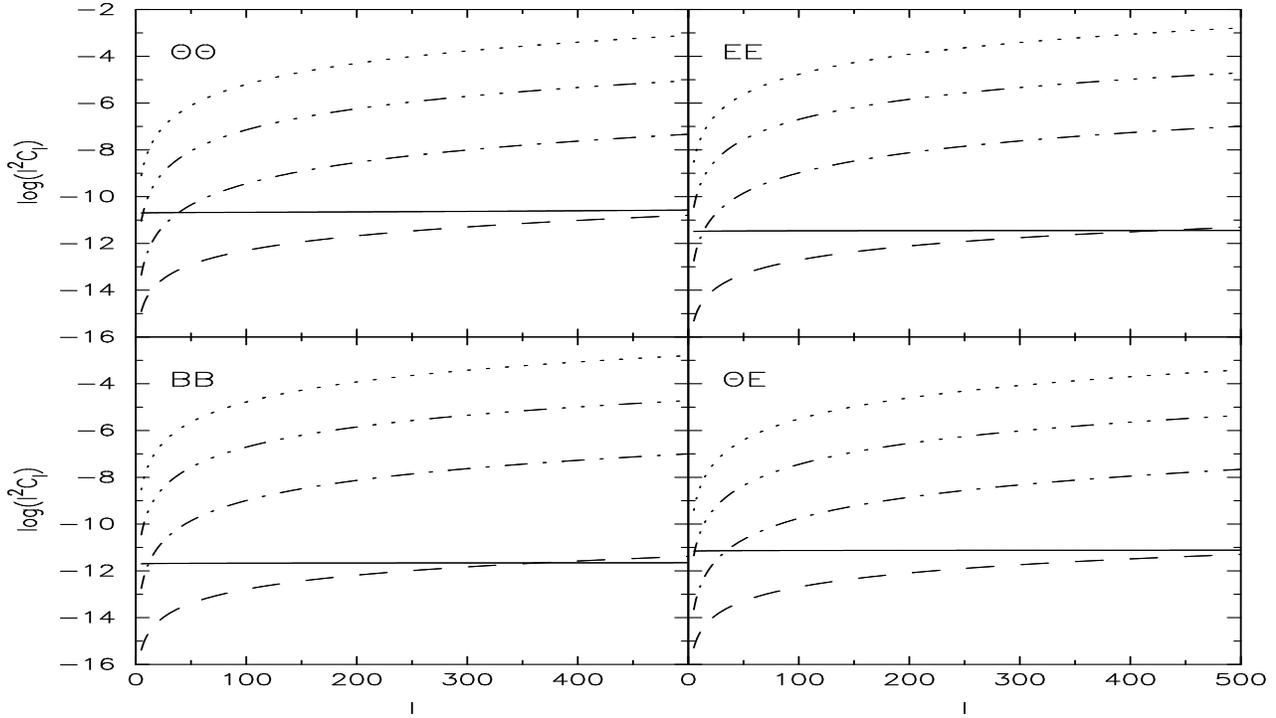}
\vskip 0.5cm
\caption{Each panel shows a single power spectrum for various values of $n$.
Solid line represents $n=-2.99$, dashed line represents $n=-2.0$,
dash-dot line represents $n=0.0$, dash-dot-dot-dot line represents $n=1.0$,
and dotted line represents $n=2.0$.}
\label{fig:Figure3}
\end{figure}

The CMB power spectra generated by a stochastic magnetic field are
plotted for $l=5$ to $l=500$ in Figs.~\ref{fig:Figure1}, \ref{fig:Figure2},
and \ref{fig:Figure3}.  Since we are interested in the signatures of the
various microwave background power spectra arising from primordial fields
that are large enough to result in the observed galactic fields via adiabatic
compression, for each plot, we choose the magnetic comoving mean-field
amplitude to be $B_\lambda=10^{-9}\,{\rm G}$ and fix $\lambda=1\,{\rm Mpc}$,
i.e. galaxy and cluster scales.  For simplicity, we consider a standard
Cold Dark Matter Universe (sCDM), i.e. a flat Universe composed of only dust
and radiation ($\eta_0\simeq6000h^{-1}\,{\rm Mpc}$) with
$\Omega_b=0.05$ and $h=0.5$.  Including a possible cosmological constant will
affect the scale factor evolution only relatively recently at redshift of a
few and will result in a slightly larger $\eta_0$.
The magnetic power spectrum cutoff wavenumbers
for vector and tensor perturbations are given by Eqs.~(\ref{eq:V-damping}) and
(\ref{eq:T-damping}) respectively.  Thus for $n=-1$ and $n=2$ for example,
with $B_\lambda=10^{-9}\,{\rm G}$ and $\lambda=1\,{\rm Mpc}$,
we have $k_D\simeq27.9\,{\rm Mpc^{-1}}$ and $k_D\simeq14.7\,{\rm Mpc^{-1}}$
respectively for vector perturbations; whereas for tensor perturbations,
we obtain $k_D\simeq80.9\,{\rm Mpc^{-1}}$ and $k_D\simeq27.1\,{\rm Mpc^{-1}}$
respectively.  For the tensor perturbations, we assume
$z_{\text{in}}/z_{\text{eq}}=10^{9}$ as in Ref.~\cite{durrer00}; the resulting
fluctuations, however, depend only logarithmically on $z_{\text{in}}$.
In our analysis, we do not decompose the magnetic field into a large
homogeneous component and a small fluctuating piece.  The stochastic magnetic
field then affects the stress-energy tensor and hence
the metric perturbations quadratically.  In computing the source terms of
vector and tensor perturbations [cf. Eqs.~(\ref{eq:V-corr-spectrum})
and (\ref{eq:T-corr-spectrum}) respectively], convolution of the magnetic
field couples the large and small scale modes, resulting in the cutoff scale
perturbations completely dominate the large scale modes for $n>-3/2$.  Thus
for $n>-3/2$, $k_D$ will determine the overall amplitude of the fluctuations.

Throughout the paper, we have been stating explicitly the terms that are
proportional to the normalization $A$ of the magnetic power spectrum.
Any power-law magnetic field can be specified
completely by the normalization $A$ and the spectral index $n$.  Since we
are interested in constraining the primordial magnetic field strength on
galaxy scales, we choose to fix $B_\lambda$ and $\lambda$ and determine
$A$ for each $n$ using Eq.~(\ref{eq:B-meansquare}).  If however one is
interested in the CMB power spectra with $A$ fixed, 
then each $n$ will give
a different value of $B_\lambda(n)$ via Eq.~(\ref{eq:B-meansquare}).
Either way will not affect the final constraints for the magnetic comoving
mean-field amplitude.  Indeed, we find it easier to constrain the
amplitude by keeping $B_\lambda$ fixed.

Figure \ref{fig:Figure1} shows the separate vector and tensor
contributions to the CMB power spectra for four different values of $n$.
For $n<-3/2$, the CMB power spectra do not depend on $k_D$, and
the size of the anisotropies increases as $n$ gets smaller;
a scale-invariant magnetic field with $n=-2.99$ generates the largest
anisotropies and hence it will yield the most stringent limit on the
primordial magnetic field (see also Fig.~1 of
\cite{durrer00}).  For $n>-3/2$, the CMB power spectra are $k_D$-dependent
and scale as $k^{2n+3}_D$; thus more and more stringent magnetic field
limits can be obtained as $n$ increases toward causal
values\footnote{Figure 1 of Ref.~\cite{durrer00} shows a weaker
and weaker upper bound for $B_\lambda$ as $n$ increases from $-3/2$.
This is because the authors there have inappropriately adopted
a smoothing scale smaller than the magnetic damping scale and have
employed a different value for the damping scale
(see Sec.~\ref{subsec:t_damping} for details).}.
For the vector perturbations, the BB power spectrum is slightly larger than
that of the EE's, as also pointed out in Ref.~\cite{hu97} that the vector
CMB polarization is dominated by the B-type modes; whereas the EE and
$\Theta{\rm E}$ power spectra are approximately identical.
Naively, the $\Theta{\rm E}$ cross correlation would be expected to be
larger than the polarization power spectra simply because the
temperature fluctuations are larger than the polarization
fluctuations. However, the temperature fluctuations are dominated by
the vector dipole term, which correlates poorly with the radial function 
describing E-type polarization.  Thus the $\Theta{\rm E}$
spectrum is dominated by a subdominant temperature contribution arising
from the vector quadrupole term, which then coincidentally renders
the spectrum a form approximately identical to the E-type polarization itself.
However in reality, the $\Theta{\rm E}$ spectrum can be slightly larger than
the polarization spectra since our approximation is good to within a
factor of three and tends to underestimate in general
(see Sec.~\ref{subsec:v_cross}).  Note that while $n\rightarrow -3$
corresponds to a scale-invariant magnetic field, the vector power
spectrum is not flat for this value. The reason is that the vorticity,
Eq.~(\ref{eq:V-vorticity-spectrum}), has an extra factor of $k$ compared 
to the magnetic field itself. The vector cross correlation 
is always negative; its absolute value is plotted.

For the tensor perturbations, the E-type is slightly larger than the B-type
polarization power spectrum.  The polarization power spectra are actually
comparable to the temperature power spectrum for $n>-3/2$.  This is due to
the additional logarithmic dependence on the magnetic damping cutoff
wavenumber for the polarization power spectra
[cf. Eqs.~(\ref{eq:T-E-power-spectrum-A}) and (\ref{eq:T-B-power-spectrum-A})],
and also because both the temperature and polarization fluctuations
are due to the intrinsic temperature quadrupole moments, which arise from
the gravitational wave solution $\dot{h}$ of 
Eq.~(\ref{eq:T-rms-metric-fluctuations}) instead of being generated via
free streaming the dipoles as in the case of the vector perturbations.
Also for $n>-3/2$, the gravitational wave source term is approximately
independent of $k$ and the resulting power spectra then possess the well-known
behavior of a white noise source $l^2C_l\propto l^3$.  Furthermore, since
tensor perturbations are damped on scales smaller than that of the vector
perturbations as discussed in Sec.~\ref{subsec:t_damping},
their induced anisotropies will then be larger than that of the vector's
for $n>-3/2$ where $k_D$ determines the overall amplitude of the microwave
background power spectra.  As expected, the tensor power spectrum
is flat for $n \rightarrow -3$ since we have a scale-invariant
magnetic field for this value.  
The tensor cross correlation is always positive.

The difference between the sign of the vector and tensor 
cross correlations can be understood from the geometric properties 
of the projection of their corresponding temperature 
and polarization sources as anisotropies on the 
sky \cite{hu97}.  The sign of the vector and tensor cross correlations is
determined by respectively (cf.\ Eq.~(80) of Ref.~\cite{hu97})
\begin{eqnarray}
\text{sgn}[C^{\Theta E(V)}_l] &=& -\text{sgn}[P^{(V)}(\dot{\tau}P^{(V)})],
\label{eq:V-cross-sign}\\
\text{sgn}[C^{\Theta E(T)}_l] &=& \text{sgn}[P^{(T)}(\dot{\tau}P^{(T)}
-\dot{h})].
\label{eq:T-cross-sign}
\end{eqnarray}
The sign of the vector cross correlation is therefore always opposite to 
that of the tensor cross correlation.

Each panel in Fig.~\ref{fig:Figure2} shows the total vector plus tensor 
contributions for the various power spectra for a particular value of $n$.
Each panel in Fig.~\ref{fig:Figure3} replots one
of the four power spectra for a range of spectral indices $n$.  As the
spectral index becomes greater than zero, the amplitudes become quite
large.  Again, this is because for $n>-3/2$, the magnetic cutoff wavenumber
$k_D$ determines the overall amplitude of the power spectra.
For a scale-invariant $n\rightarrow -3$ spectral index, a magnetic
comoving mean-field amplitude of $B_\lambda=10^{-9}\,{\rm G}$, and a
comoving smoothing scale of $\lambda=1\,{\rm Mpc}$, at $l=500$ for example,
the temperature and polarization power spectra are smaller than
the amplitudes as expected from scale-invariant density perturbations
normalized to COBE (i.e., CMB fluctuations in ``standard'' cosmological
models).  But for $n=0$, the temperature power spectrum is essentially a
factor of 50 whereas the E-type polarization power spectrum 
is a factor of $10^{4}$ larger than that expected from the scalar sCDM
model at $l=500$.  Therefore, observational limits will be much
stronger for causal fields than for scale-invariant fields.

To estimate the potential observational limits on stochastic magnetic
fields, the induced microwave background anisotropies must be large
enough to be disentangled from the anisotropies arising from density
perturbations. Current temperature maps give power spectrum
measurements with error bars on the order of 10\% out to $l=400$ for
bins of width $\Delta l = 50$ \cite{hanany00,bernardis00}. The MAP
satellite, which is already in orbit, will make temperature
measurements out to around $l=800$ and will reach the cosmic variance
limit, $\Delta C_l = (l+1/2)^{-1/2} C_l$, out to $l=400$ \cite{map}.
By the end of the decade, and perhaps within five years, we can expect
a cosmic-variance limited temperature power spectrum measurement to
$l=3000$. Polarization fluctuations will also be detected soon, and
the progress in their measurement will likely lag temperature
fluctuations by about a decade.  A rough but conservative estimate is
that a magnetic-field signal which is at least 10\% of the dominant
density-perturbation signal will be detectable. The ultimate
sensitivity in measuring the temperature power spectrum at, say,
$l=500$ will be significantly better than this, and the extent to
which magnetic fields can be constrained depends more on the
degeneracy of the magnetic field signal with shifts in various
cosmological parameters. Basic statistical techniques for pursuing
such an analysis are well-known (see, e.g., \cite{kamionkowski98}) and
will be considered elsewhere.

Using this crude 10\% criterion, we can anticipate constraints on
stochastic magnetic fields from upcoming temperature measurements
(e.g., the MAP satellite, currently taking data) by simply comparing
the predicted amplitude at $l=500$ to the amplitude of current
measurements, which is on the order of $l^2 C_l \simeq 10^{-9}$.  We
assume that the remainder of the power spectrum is used for
discrimination between the signals from magnetic fields and other
temperature power spectrum contributors.  For the scale-invariant
magnetic field with $n\rightarrow -3$, a comoving mean-field amplitude
of $B_\lambda=10^{-9}\,{\rm G}$ gives temperature anisotropies at the
level of $\sim3\%$ of current measurements.  Since for $n<-3/2$,
$l^2C_l\propto B^4_\lambda$, the constraint from temperature
perturbations on a comoving $1\,{\rm Mpc}$ scale will be around
$1.4\times10^{-9}\,{\rm G}$, which is approximately at the same level
as the previous constraints for a primordial homogeneous magnetic
field \cite{barrow97,adams96,durrer98}.  The addition of E-type
polarization measurements here will improve the constraint, since the
ratio of the E-type polarization to temperature power spectra is
larger for stochastic magnetic fields than the dominant density
perturbations.  For $n>-3/2$, the polarization power spectra are
comparable to the temperature power spectrum due to the dominant
tensor perturbations; thus E-type polarization measurements will yield
more stringent constraints than temperature measurements alone.  Here
we will be conservative and project stochastic magnetic field
constraints using temperature measurements only.  For $n>-3/2$, we
have $l^2C_l\propto B^{14/(n+5)}_\lambda$, where
$14/(n+5)=4+[-2/(n+5)](2n+3)$, since $l^2C_l\propto A^2k^{2n+3}_D$,
$A\propto B^2_\lambda$ [cf. Eq.~(\ref{eq:B-meansquare})], and
$k_D\propto B^{-2/(n+5)}_\lambda$ [cf. Eqs.~(\ref{eq:V-damping}) and
(\ref{eq:T-damping})].  As $n$ increases towards causal values, the
amplitude of the temperature fluctuations increases as $k^{2n+3}_D$
and hence the constraints become stronger.  At $n=0$, $l^2 C_l$ at
$l=500$ is approximately $5\times10^{-8}$, which will yield a
constraint on $B_\lambda$ of
$\left(\frac{5\times10^{-8}}{0.1\times10^{-9}}\right)^{-5/14}
\times10^{-9}\,{\rm G}\simeq10^{-10}\,{\rm G}$.  For the causal field
$n=2$, the constraint on $B_\lambda$ will be as small as
$4\times10^{-13}\,{\rm G}$.  Such constraints will be stronger than
any current limits on Mpc-scale primordial stochastic magnetic fields
at decoupling.

Ultimately, B-type polarization has the greatest potential for
constraining primordial magnetic fields. This is a cleaner signature,
because primordial scalar (density) perturbations produce none
\cite{kamionkowski97a,zaldarriaga97}.  Aside from polarized foreground
emission, the only other expected sources are from primordial tensor
perturbations and from gravitational lensing \cite{zaldarriaga98}. Tensor
perturbations with a spectrum near scale-invariant will give
significant anisotropies only at large angular scales ($l<100$), while
lensing contributes mainly at small angular scales ($l>500$).
Stochastic magnetic fields will contribute on intermediate scales and
should be clearly distinguishable. If foreground emission can be
separated from its frequency dependence, limits on $B_\lambda$ from B-type
polarization should be determined purely by measurement error bars on
$C_l^{BB}$. Note that a primordial
magnetic field also generates an additional B-type
polarization signal via Faraday rotation of the CMB polarization
\cite{kosowsky96}.  This signal will be negligible compared to the
direct B-type polarization signal for any frequency of practical interest.

All of the results in this paper have been obtained via analytic
approximations to the exact solutions.  Apart from the vector
cross-correlation power spectrum, the accuracy of the results is as good
as the quality of the analytic approximations to various expressions, except
that the vector temperature case has neglected an additional
few percent temperature contribution arising from the angular dependence of
polarization and the vector potential.  These approximations are all 
discussed in the text; in sum, they are good to within 20\% over the
range of parameters considered, with the exception of the vector
temperature case in the regime $-2.5\leq n<-2$,
which is good to within 30\%.  Meanwhile,
accuracy of the vector cross-correlation power spectrum is only good to
within a factor of three, since we have neglected the two terms in
Eq.~(\ref{eq:V-cross-power-spectrum}) arising from the correlation between
the temperature dipole and the E-type polarization radial functions.
It is important to realize that errors in these analytic
approximations will have negligible effects on the estimation of the
magnetic field limits for $n\leq2$ since the amplitude of each power spectrum
scales as $B^4_\lambda$ for $n<-3/2$ and $B^{14/(n+5)}_\lambda$
for $n>-3/2$.

In this paper, we have focussed on the magnetic field-induced 
microwave background anisotropies
for $l\leq500$, where the analysis is relatively simple and free
from the detailed microphysics of recombination. 
We have only considered
vector and tensor metric perturbations; for smaller angular
scales $500<l<2000$, the magnetic-induced CMB anisotropies are dominated
by vector perturbations \cite{seshadri01}.  Stochastic magnetic fields will
also produce scalar perturbations. This case is significantly more
complex due to physical compensation effects and the large number of
terms involved in the relevant expressions. Rough
analytic estimates show that including the scalar
results will only modestly improve the magnetic field constraints given in this
paper, since radiation pressure prevents the induced density fluctuations from
growing effectively before recombination and the compressional modes
are erased up to the Silk scale $L_S$ \cite{jedamzik98,subramanian98a}. 
Therefore scalar perturbations will generally give a subdominant
contribution to the microwave background anisotropy.

Our results suggest that while it may be plausible for primordial stochastic
fields with $n\alt0$ to result in the observed galactic fields via adiabatic
compression alone, it will be very difficult for causal fields without
invoking some form of dynamo mechanism.  In a recent paper \cite{caprini02},
a similar calculation using the nucleosynthesis bound on gravitational
radiation induced by the anisotropic stress of a primordial stochastic
magnetic field yields extremely stringent limits on the galactic-scale
magnetic field amplitudes ($B_\lambda\leq10^{-27}\,{\rm G}$) for fields
generated at the electroweak phase transition or earlier, thus ruling out
most of the magnetogenesis processes for primordial fields seeding the
observed large-scale coherent galactic fields.

\acknowledgments

We are extremely grateful to R.~Durrer for numerous explanations and
invaluable comments and suggestions.  W.~Hu and M.~White contributed other
helpful discussions.  The referee, Karsten Jedamzik,
alerted us to inadequate treatments of vorticity and magnetic
damping in an earlier version of this paper, in addition to
useful discussions during its initial preparation.  This work has been
supported by the COBASE program of the U.S.\ National Research Council
and by the NASA Astrophysics Theory Program through grant NAG5-7015.
A.K.\ is a Cotrell Scholar of the Research Corporation.  T.K.\
acknowledges the kind hospitality of Rutgers University.

\appendix
\section{Derivation of the Vector Isotropic Spectrum}
\label{sec:vector_derive}

Our objective is to derive the vector isotropic spectrum
$|\Pi^{(V)}(k)|^2$ defined in Eq.~(\ref{eq:2pt-fcn-vector1}), which
will be useful for calculating vector CMB power spectra.  Using
Eq.~(\ref{eq:B-vector}), the two-point correlation function of
$\Pi^{(V)}_i$ is given by
\begin{equation}
\langle\Pi^{(V)}_i({\mathbf k})\Pi^{(V)*}_i({\mathbf k'})
\rangle=P_{ib}\hat{k}_aP_{id}'\hat{k}_c'\langle\tau^{(B)}_{ab}
({\mathbf k})\tau^{(B)*}_{cd}({\mathbf k'})\rangle,
\label{eq:2pt-fcn-vector2}
\end{equation}
where $P_{id}'=\delta_{id}-\hat{k}_i'\hat{k}_d'$.  
We simplify our calculation by splitting the electromagnetic
stress-energy tensor into two pieces: 
$\tau^{(B)}_{ij}({\mathbf k})=\tau^{(B,1)}_{ij}({\mathbf k})
+\tau^{(B,2)}_{ij}({\mathbf k})$ where
\begin{mathletters}
\label{eq:Bk-stress-1-2}
\begin{eqnarray}
\tau^{(B,1)}_{ij}({\mathbf k}) 
&\equiv& \frac{1}{(2\pi)^3}\frac{1}{4\pi}
\int d^3p\,B_i({\mathbf
p})B_j({\mathbf k-p}),
\label{eq:Bk-stress-1-2a}\\
\tau^{(B,2)}_{ij}({\mathbf k}) 
&\equiv& -\frac{1}{(2\pi)^3}\frac{1}{8\pi}
\delta_{ij}\int d^3p\,B_l({\mathbf
p})B_l({\mathbf k-p}).
\label{eq:Bk-stress-1-2b}
\end{eqnarray}
\end{mathletters}
The two-point correlation function of the electromagnetic
stress-energy tensor in Eq.~(\ref{eq:2pt-fcn-vector2}) will now be
described by a sum of four two-point correlation functions:
\begin{eqnarray}
\langle\tau^{(B)}_{ab}({\mathbf k})\tau^{(B)*}_{cd}({\mathbf k'})\rangle 
&=& \langle\tau^{(B,1)}_{ab}({\mathbf k})
\tau^{(B,1)*}_{cd}({\mathbf k'})\rangle
+\langle\tau^{(B,1)}_{ab}({\mathbf k})
\tau^{(B,2)*}_{cd}({\mathbf k'})\rangle\nonumber\\
& & \mbox{} +\langle\tau^{(B,2)}_{ab}({\mathbf k})\tau^{(B,1)*}_{cd}
({\mathbf k'})\rangle+\langle\tau^{(B,2)}_{ab}({\mathbf k})
\tau^{(B,2)*}_{cd}({\mathbf k'})\rangle.
\label{eq:2pt-fcn-Bkstress}
\end{eqnarray}
Only $\langle\tau^{(B,1)}_{ab}\tau^{(B,1)*}_{cd}\rangle$ above has a
non-vanishing contribution toward the two-point correlation function
of $\Pi^{(V)}_i$ in Eq.~(\ref{eq:2pt-fcn-vector2}), since each of the
remaining correlation functions in Eq.~(\ref{eq:2pt-fcn-Bkstress})
contains either $\delta_{ab}$, $\delta_{cd}$, or both, and will vanish
when they are acted upon by $P_{ib}\hat{k}_aP_{id}'\hat{k}_c'$.  We
can now rewrite Eq.~(\ref{eq:2pt-fcn-vector2}) as
\begin{equation}
\langle\Pi^{(V)}_i({\mathbf k})\Pi^{(V)*}_i({\mathbf k'})\rangle
=P_{ib}\hat{k}_aP_{id}'\hat{k}_c'\langle\tau^{(B,1)}_{ab}
({\mathbf k})\tau^{(B,1)*}_{cd}({\mathbf k'})\rangle.
\label{eq:2pt-fcn-vector3}
\end{equation}
We can evaluate the two-point correlation function
$\langle\tau^{(B,1)}_{ab}\tau^{(B,1)*}_{cd}\rangle$ as follows.
Beginning with the definition of Eq.~(\ref{eq:Bk-stress-1-2a}), we
assume the random magnetic field is Gaussian and apply Wick's theorem
\begin{eqnarray}
\langle B_i({\mathbf k}_i)B_j({\mathbf k}_j)
B_l({\mathbf k}_l)B_m({\mathbf k}_m)\rangle 
&=& \langle B_i({\mathbf k}_i)B_j({\mathbf k}_j)\rangle
\langle B_l({\mathbf k}_l)B_m({\mathbf k}_m)\rangle
+\langle B_i({\mathbf k}_i)B_l({\mathbf k}_l)\rangle
\langle B_j({\mathbf k}_j)B_m({\mathbf k}_m)\rangle \nonumber \\
& & \mbox{} +\langle B_i({\mathbf k}_i)B_m({\mathbf k}_m)
\rangle\langle B_j({\mathbf k}_j)B_l({\mathbf k}_l)\rangle
\label{eq:Wick}
\end{eqnarray}
and the reality condition $B^{*}_i({\mathbf k})=B_i({\mathbf -k})$,
and finally use Eq.~(\ref{eq:2pt-fcn-form}) for the form of the
stochastic magnetic field two-point correlation function
to arrive at (see also \cite{durrer00})
\begin{eqnarray}
\langle\tau^{(B,1)}_{ab}({\mathbf k})\tau^{(B,1)*}_{cd}({\mathbf k'})\rangle 
&=& \frac{1}{(4\pi)^2}\int d^3p\,P(p)P(|{\mathbf k-p}|)
[(\delta_{ac}-\hat{p}_a\hat{p}_c)(\delta_{bd}
-(\widehat{{\mathbf k-p}})_b(\widehat{{\mathbf k-p}})_d) \nonumber \\
& & \mbox{} +(\delta_{ad}-\hat{p}_a\hat{p}_d)(\delta_{bc}
-(\widehat{{\mathbf k-p}})_b(\widehat{{\mathbf k-p}})_c)]
\delta({\mathbf k-k'}).
\label{eq:Bk1-2pt-1}
\end{eqnarray}
Substitute Eq.~(\ref{eq:Bk1-2pt-1}) into
Eq.~(\ref{eq:2pt-fcn-vector3}) and define 
$\gamma\equiv{\mathbf\hat{k} \cdot\hat{p}}$, 
$\beta\equiv{\mathbf\hat{k}\cdot(\widehat{k-p})}$, 
and $\mu\equiv{\mathbf\hat{p}\cdot(\widehat{k-p})}$
to obtain the two-point correlation function of the vector $\Pi^{(V)}_i$: 
\begin{equation}
\langle\Pi^{(V)}_i({\mathbf k})\Pi^{(V)*}_i({\mathbf k'})
\rangle=\frac{1}{(4\pi)^2}
\int d^3p\,P(p)P(|{\mathbf k-p}|)[(1-\gamma^2)(1+\beta^2)
+\gamma\beta(\mu-\gamma\beta)]\delta({\mathbf k-k'}).
\label{eq:2pt-fcn-vector4}
\end{equation}
The integral above is similar to the mode-coupling integral $I^2(k)$
in Eq.~(11) of Ref.~\cite{subramanian98b}.  Although it cannot be evaluated
analytically, terms within the square bracket are products of cosine
factors; hence the square bracket itself can be approximated by unity,
which has essentially been done in Ref.~\cite{durrer00}.  Comparing with
Eq.~(\ref{eq:2pt-fcn-vector1}) gives
\begin{equation}
|\Pi^{(V)}(k)|^2\simeq\frac{1}{8(2\pi)^2}
\int d^3p\,P(p)P(|{\mathbf k-p}|).
\label{eq:V-corr-spectrum1}
\end{equation}
Using the expression for $P(k)$ in Sec.~\ref{sec:spectrum} and
choosing $\hat{\mathbf k}$ to be the polar axis, the vector isotropic
spectrum becomes
\begin{equation}
|\Pi^{(V)}(k)|^2\simeq\frac{(2\pi)^{2n+9}}{32}\frac{B^4_\lambda}
{\Gamma^2\left(\frac{n+3}{2}\right)k^{2n+6}_\lambda}
\int^{k_D}_0 dp\,p^{n+2}\int^{1}_{-1}d\gamma\, 
(k^2+p^2-2kp\gamma)^{n/2}.
\label{eq:V-corr-spectrum2}
\end{equation}
The integral over $\gamma$ is
\begin{equation}
\int^{1}_{-1}d\gamma\,(k^2+p^2-2kp\gamma)^{n/2}
=\frac{1}{kp(n+2)}[(k+p)^{n+2}-|k-p|^{n+2}],
\label{eq:gamma-integral}
\end{equation}
and the expression within the square bracket above can be approximated as
\begin{equation}
(k+p)^{n+2}-|k-p|^{n+2}\simeq\cases{2(n+2)k^{n+1}p, & $p<k$;\cr
                                    2(n+2)kp^{n+1}, & otherwise.\cr}
\label{eq:k-p-relation}
\end{equation}
Substituting Eqs.~(\ref{eq:gamma-integral}) and
(\ref{eq:k-p-relation}) into Eq.~(\ref{eq:V-corr-spectrum2}) and
evaluating, we finally arrive at the expression for the vector
isotropic spectrum, Eq.~(\ref{eq:V-corr-spectrum}).

\end{document}